\documentclass[12pt]{iopart}

\usepackage{graphicx}
\usepackage{dcolumn}
\usepackage{bm}
\usepackage{amssymb,amsfonts}
\usepackage[utf8]{inputenc}
\usepackage[T1]{fontenc}
\usepackage{mathptmx}
\usepackage{enumerate}   
\usepackage[normalem]{ulem}
\DeclareSymbolFont{greekletters}{OML}{cmr}{m}{it}
\DeclareMathSymbol{\varrho}{\mathord}{greekletters}{"25}

\usepackage{cite}
\usepackage{xcolor}

\begin{document}

\title[Explicit noise and dissipation operators for quantum stochastic thermodynamics]{Explicit noise and dissipation operators for quantum stochastic thermodynamics}

\author{Stefano Giordano}%
\address{University of Lille, CNRS, Centrale Lille, Univ. Polytechnique Hauts-de-France, UMR 8520 - IEMN - Institut d'{\'E}lectronique, de Micro{\'e}lectronique et de Nanotechnologie, F-59000 Lille, France}
\ead{stefano.giordano@univ-lille.fr}  

\author{Fabrizio Cleri}%
\address{University of Lille, Institut d'{\'E}lectronique, de Micro{\'e}lectronique et de Nanotechnologie (IEMN CNRS UMR8520) and Departement de Physique, F-59652 Villeneuve d'Ascq, France}
\ead{fabrizio.cleri@univ-lille.fr}  

\author{Ralf Blossey}
\address{University of Lille, Unit{\'e} de Glycobiologie Structurale et Fonctionnelle (UGSF), CNRS UMR8576, F-59000 Lille, France}
\ead{ralf.blossey@univ-lille.fr}

\begin{abstract}
The theory of open quantum systems plays a fundamental role in 
several scientific and technological disciplines, from quantum computing and information science to molecular electronics and quantum thermodynamics. Despite its widespread 
relevance, a rigorous formulation of quantum dissipation 
in conjunction with thermal noise remains a topic of active research. In this work, we establish a formal correspondence between classical stochastic thermodynamics, in particular the Fokker-Planck and Klein-Kramers equations, and the quantum master equation. Building on prior studies of multiplicative noise in classical stochastic differential equations, we demonstrate that thermal noise at the quantum level manifests as a multidimensional geometric stochastic process. By applying canonical quantization, we introduce a novel Hermitian dissipation operator that serves as a quantum analogue of classical viscous friction. This operator allows for a well-defined expression of heat exchange between a system and its environment, enabling the formulation of an alternative quantum equipartition theorem. Our framework ensures a precise energy balance that aligns with the first law of thermodynamics and an entropy balance consistent with the second law. The theoretical formalism is applied to two prototypical quantum systems, 
the harmonic oscillator and a particle in an infinite potential well, 
for which it provides
new insights into nonequilibrium thermodynamics at the quantum scale. 
Our results advance the understanding of dissipation in quantum systems and establish a foundation for future studies on stochastic thermodynamics in the quantum domain.
\end{abstract}

%
%
\submitto{\JSTAT}
%
%
%

\newpage
\section{Introduction}

The study of open quantum systems emerged from the realization that quantum systems rarely exist in perfect isolation. Instead, they are often coupled to external environments, which introduce noise and dissipation. The interaction of a system with its surroundings causes energy exchange and decoherence, influencing its behavior.
One of the earliest approaches, introduced in the 1940s, incorporated nonconservative forces into quantum mechanics within Schrödinger's framework, ultimately leading to the renowned Caldirola-Kanai equation \cite{caldirola1941,kanai1948}.
This approach is based on the fact that the presence of nonconservative forces in otherwise Hamiltonian systems can be studied through fully Hamiltonian systems by means of a nonlinear time transformation (this is a method introduced by Levi-Civita) \cite{caldirola1941,kanai1948}. 
Despite facing significant criticism, this approach was shown to be consistent with the physics of open systems, paving the way for the development of this new direction in quantum theory \cite{caldirola1982,caldirola1983}. 
On this basis, quantum dissipation has been investigated extensively \cite{svinin1975,stocker1979,dekker1981}.

In the following years, the field advanced with the introduction of the quantum master equation, which describes the evolution of an open system’s density matrix \cite{redfield1965,blum1981,lindenberg1990,zwanzig2001}. 
One of the first examples is given by the Redfield master equation, which describes the time evolution of a quantum system  weakly coupled to the environment \cite{redfield1965}. It was used in nuclear magnetic resonance spectroscopy 
(NMR), but it does not always guarantee a positive time evolution of the density matrix \cite{trushechkin2021}.
These studies opened the way to various approaches based on the path integral formulation, the foundations of quantum thermodynamics, and quantum Brownian motion \cite{caldeira1981,caldeira1983,caldeira1983bis,beretta1984,beretta1985,jannussis1985,ohba1997,cavalcanti1998,ghosh2024}.

Simultaneously, starting in the 1970s, a more sophisticated line of research aimed at finding the most complete mathematical formulation to describe the evolution of the density matrix of an open system. This effort culminated in the development of the Gorini-Kossakowski-Sudarshan-Lindblad (GKLS) equation, which characterizes the generator of a quantum dynamical semigroup \cite{lindblad1974,lindblad1975,gorini1976,lindblad1976,pascazio2017}.
This result represents a Markovian evolution that refines the approximations introduced earlier by the Redfield equation \cite{redfield1965}, 
and
provided a significant boost to the rigorous advancement of quantum thermodynamics \cite{lebowitz1978,trushechkin2018,trushechkin2019,ruelle2001,pillet2002,pucci2013,kosloff2019,prior2010,pigeon2016,campaioli2024}.
These theories led to a deeper understanding of how coherence is lost in quantum systems and the role of the environment in quantum measurements \cite{schlosshauer2004,toth2014,schlosshauer2019}.
As the field matured, researchers developed more refined models to describe quantum equipartition theorems \cite{bialas2018,spiechowicz2018,bialas2019,luczka2020,tong2024}, and quantum fluctuation theorems \cite{tal1,tal2,tal2bis,tal3}. 
Recently, a critical review of the assumptions and limitations of Lindblad's quantum master equation has been carried out \cite{stefanini}.

The theory of open quantum systems has established itself as a fundamental framework in various scientific and technological domains. Its applications span nuclear magnetic resonance \cite{abragam1961,slichter1990}, optical pumping \cite{happer1972}, masers \cite{scovil1959,thomas2020}, and lasers \cite{youssef2009,dorfman2018}, as well as quantum computing \cite{deutsch1985,buluta2009,mermin2012,cleri2024}, quantum information \cite{vedral2006,hayashi2017} and molecular electronics \cite{ke2008,jensen2022}. Moreover, it provides critical insights into emerging fields such as quantum biology \cite{landsberg1984,mohseni2014,bianco1992,chiabrera2000}, including photosynthesis \cite{engel2007,sarovar2010,scholes2011}, and plays a pivotal role in the study of quantum thermodynamic devices \cite{myers2022} and the understanding of the arrow of time in the transition between classical and quantum thermodynamics \cite{maddox1985,lebowitz1993,guff2025}.
Additionally, the framework of open quantum systems has profound implications in emerging fields such as quantum metrology and sensing, where controlled decoherence enhances precision in devices like atomic clocks and quantum interferometers \cite{giovannetti2011,pezze2018,braun2018}. It also plays a critical role in optomechanics, enabling the study of light-matter interactions at the quantum level with applications in gravitational wave detection \cite{barzanjeh2022,huang2024}. Moreover, open quantum systems are integral in understanding the dynamics of complex out-of-equilibrium systems, such as those observed in nanoscale materials and structures \cite{keimer2017,tokura2017,alfieri2023,goyal2025}.

It is interesting to compare the
quantum approaches
with parallel developments in classical stochastic thermodynamics. 
In this line of research, building on Langevin's foundational work \cite{langevin}, the Fokker-Planck equation was initially formulated for a generic stochastic system \cite{fokker,planck}, while the Klein-Kramers equation was developed for a mechanical system studied in phase space \cite{klein,kramers}. 
Stochastic thermodynamics in its true form was later introduced with the definition of heat for a single random trajectory \cite{seki1,seki2}, leading to a microscopic formulation of the first law. 
Subsequently, the concept of entropy was also defined for individual system trajectories \cite{sei1,sei2,sei3}, allowing the second law to be associated with Langevin's stochastic approach.
Today, the theory of stochastic processes based on Langevin and Fokker-Planck equations is well-established and widely applied across various fields of physics, chemistry, and engineering \cite{vankampen,risken,coffey,schuss1,schuss2}. 
Moreover, the laws of thermodynamics derived from a microscopic stochastic foundation have been extensively studied and applied to numerous physical systems \cite{sch,espo1,espo2,tome0,tome1,tome2}, including holonomic underdamped and overdamped systems \cite{pan,mura,annphys,giorda}.

In this work, we aim to explore in detail the correspondences between the classical Fokker-Planck or Klein-Kramers equations and the quantum master equation. 
To achieve this, we first examine the concept of thermal noise in quantum mechanics, building on our previous studies, where we investigated the effects of multiplicative noise in stochastic differential equations \cite{palla2020,geom,dupont2024,giordano2024}. 
Specifically, we demonstrate how thermal noise manifests 
itself at the quantum level as a multidimensional geometric stochastic process, extending some of the results obtained in Ref. \cite{geom}.
We then introduce quantum dissipation by leveraging the analogy with classical equations, specifically applying canonical quantization, which replaces Poisson brackets with commutators \cite{landauq,cohen}. The search for analogies between the evolution equations of classical and quantum open systems is not new, and 
several studies have explored this direction \cite{bianco,oliveira2016,oliveira2023,oliveira2024}.
The original contribution of our work is the definition of a Hermitian dissipation or friction operator, which corresponds to a physical observable and replaces the classical momentum in describing the viscous friction typical of the Langevin approach. 
When expressed in the energy basis of the system, this operator closely resembles the momentum operator, making the physical interpretation of the proposed model more intuitive. 
The fact that this operator is Hermitian also allows for the construction of a mathematically well-defined expression for the heat exchanged between the system and the external environment, providing a clear physical interpretation. 
This, in turn, enables the formulation of an alternative quantum equipartition theorem compared to those proposed in the existing literature \cite{bialas2018,spiechowicz2018,bialas2019,luczka2020,tong2024}.
The explicit form of the heat exchanged between the system and the environment allows for an energy balance that precisely corresponds to the first law of thermodynamics. 
Similarly, we can develop an entropy balance which also leads to the second law.
In the latter, we can identify the entropy flow rate, associated with heat transfer between the system and the environment, and the entropy production rate, which corresponds to the irreversibility of the thermodynamic relaxation process. 
The entropy production is always positive to satisfy the second law \cite{lebowitz1978,trushechkin2018,trushechkin2019,ruelle2001,pillet2002,kosloff2019}.
This law is also discussed in terms of the free energy, a quantity that always decreases during the evolution toward thermodynamic equilibrium.
 We present two alternative integral
expressions for 
the Hermitian friction operator, both valid regardless of the chosen basis, along with an explicit form in the energy basis.

Finally, we have implemented the mathematical apparatus to describe the nonequilibrium dynamics of both the harmonic oscillator and a particle in an infinite potential well,
providing a comprehensive understanding of the thermodynamic behavior of these ubiquitously used model systems.

\section{Classical approach to stochastic thermodynamics}

In order to develop our approach to the noisy quantum dynamics in parallel to the notions of classical stochastic thermodynamics, we begin by
recalling the key results in this domain.
We consider a classical system composed of $N$ particles with masses $m_i$, described by coordinates $\vec r_i$, velocities
$ \vec v_i =  \frac{d}{dt} \vec r_i$, and linear momenta $\vec p_i=m_i \vec v_i$ $(i=1,...,N)$.
Newton's equations of motion
are given by $\vec F_i   =  m_i \vec a_i $, where $\vec F_i$ is the total force acting on the particle, and $\vec a_i=\frac{d}{dt} \vec v_i$ is the acceleration vector.
The total kinetic energy of the system $K_0$ can be written as $K_0 = \sum_{i=1}^{N} \frac{m_i}{2} \vec v_i \cdot \vec v_i$. 

We suppose that the total force
$\vec F_i$ includes the following contributions: (i) a conservative force field describing the system structure, (ii) an external force field representing the work done on the system, (iii) a friction force mimicking the energy transfer from the particles to the thermal bath, and (iv) a noise term mimicking the energy transfer from the bath to the system. The friction and noise forces (iii) and (iv) represent the so-called Langevin thermal bath \cite{langevin}.
We then postulate that
\begin{equation}\fl
 \label{totforce}
 \vec F_i =- \frac{\partial V_0}{\partial \vec r_i} +\vec{f}_i(t)-m_i \beta \vec v_i + \sqrt{D m_i} \vec n_i(t) , 
\end{equation}
where $\beta$ is the friction coefficient (per unit mass) and $D$ is the diffusion coefficient (per unit mass). As usual, we assume the following hypotheses on the noises: $\vec n_i(t) \in \mathbb{R}^3$ are Gaussian stochastic processes with expectation value $\mathbb{E}\{ \vec n_i(t)\} = 0$, and correlation $\mathbb{E}\{\vec n_i(t_1) \otimes \vec n_j(t_2)\} = 2 \delta_{ij}{I}_3 \delta(t_1-t_2)$. 
Here, $\delta_{ij}$ is the Kronecker delta, $\delta(.)$ is the Dirac delta function, $\otimes$ is the tensor product of vectors, and ${I}_3$ is the $3\times3$ identity matrix \cite{vankampen,risken,coffey}. 
This formulation
is consistent with the canonical distribution in equilibrium, and with the first and second laws of thermodynamics during the out-of-equilibrium evolution, as we briefly discussed below. 
The system dynamics can also be stated in terms of the Hamilton equations 
\begin{eqnarray}
\label{ham}
\fl {\dot {\vec r}}_i = \frac{1}{m_i}{\vec p}_i, \,\,\,\,\,{\dot {\vec p}}_i = - \frac{\partial V_0}{\partial \vec r_i} +\vec{f}_i(t)-\beta \vec p_i + \sqrt{D m_i} \vec n_i(t).
\end{eqnarray}
From a mathematical point of view, 
Eq.(\ref{ham}) represents a stochastic differential problem with additive noise \cite{schuss1,schuss2}.
We can now apply the Fokker-Planck methodology, which is briefly presented here for an arbitrary interpretation of the stochastic calculus. Although this distinction is strictly speaking not relevant here since the classical Hamilton equations have an additive noise, we will see that the corresponding quantum equations have multiplicative noise, where this distinction will be essential \cite{vankampen,risken,coffey}. We consider the stochastic differential system
\begin{eqnarray}\fl
 \label{stoch}
 \frac{dx_i }{dt} = h_i(\vec x,t) + \sum_{j=1}^\mathfrak{m} g_{ij}(\vec x,t) n_j(t),
\end{eqnarray}
with $\mathfrak{n}$ equations and $\mathfrak{m}$ noise terms ($\forall i = 1, ..., \mathfrak{n}, \forall j = 1, ..., \mathfrak{m}$).
It assumes a precise meaning only after declaring the adopted interpretation of the stochastic calculus. In order to achieve this, we must 
specify the parameter $\alpha$, with $0\le \alpha \le 1$, that defines 
the position of the point at which we calculate any integrated function in the small intervals of the adopted Riemann sum. 
The Gaussian noises $n_j(t)$ ($\forall j = 1, ..., \mathfrak{m}$) satisfy the properties $\mathbb{E}\{  n_j(t)\} = 0$ and $\mathbb{E}\{ n_i(t_1)  n_j(t_2)\} = 2 \delta_{ij} \delta(t_1-t_2) $. 
The introduced stochastic differential equation corresponds to the following evolution equation for the probability density  $W(\vec x,t)$ (the Fokker-Planck equation) \cite{risken,coffey}
\begin{eqnarray}
\nonumber
\fl \frac{\partial W(\vec x,t)}{\partial t} &=& -  \sum_{i=1}^\mathfrak{n} \frac{\partial}{\partial x_i} [h_i W(\vec x,t)]-  \sum_{i=1}^\mathfrak{n} \frac{\partial}{\partial x_i} \left\{ 2\alpha\left[ \sum_{k=1}^\mathfrak{n} \sum_{j=1}^\mathfrak{m} g_{kj} \frac{\partial g_{ij}}{\partial x_k}   \right]  W(\vec x,t) \right\} \\  \label{fokpla}\fl
 &&+   \sum_{i=1}^\mathfrak{n} \sum_{j=1}^\mathfrak{m} \frac{\partial^2 }{\partial x_i \partial x_j} \left\{ \left[ \sum_{k=1}^\mathfrak{m} g_{ik} g_{jk}  \right]  W(\vec x,t) \right\},
\end{eqnarray}
where the first term represents the drift, the second is the noise induced drift (which depends on $\alpha$) and the third  
the diffusion (characterizing the effect of the noise). This expression includes the It\^{o} ($\alpha=0$) \cite{itofp}, the Stratonovich ($\alpha=1/2$) \cite{stratofp} and the H\"{a}nggi-Klimontovich ($\alpha=1$) \cite{hanggifp,klimofp} as particular cases (see 
Refs.\cite{sokolov,bo2019}). It is interesting to observe that the theory can be generalized to take into consideration the possible cross-correlation of the noises \cite{denisov1,denisov2}.

Eventually, we can write the Fokker-Planck \cite{fokker,planck} (or Klein-Kramers \cite{klein,kramers}) equation associated with Eqs.(\ref{ham}) in the following form 
\begin{eqnarray}
\label{klein1}\fl
 \frac{\partial W}{\partial t} =  -\sum_{i=1}^N \frac{{\vec p}_i}{m_i}\cdot \frac{\partial W}{\partial {\vec r}_i} +\sum_{i=1}^N  \frac{\partial V}{\partial {\vec r}_i} \cdot \frac{\partial W}{\partial {\vec p}_i}  
  + 3N \beta W + \beta \sum_{i=1}^N{\vec p}_i\cdot   \frac{\partial W}{\partial{\vec p}_i} + D\sum_{i=1}^N m_i \frac{\partial^2 W }{\partial {\vec p}_i ^2},
\end{eqnarray}
where $W=W({\vec r}_1,...,{\vec r}_N,{\vec p}_1,...,{\vec p}_N,t)$, and we introduced the effective potential energy $V=V_0-\sum_{i=1}^N\vec{f}_i\cdot \vec{r}_i$. The derivative $\frac{\partial^2 W }{\partial {\vec p}_i ^2}$ represents the Laplacian operator with respect the three components of ${\vec p}_i$.
A more interesting form for the following developments can be found by introducing the Poisson brackets as follows
\cite{gantmacher}
\begin{eqnarray}
\nonumber
\fl \frac{\partial W}{\partial t} &=& \left\lbrace \mathcal{H},W \right\rbrace  + \beta \sum_{i=1}^N \left( \left\lbrace x_{i},p_{xi} W\right\rbrace+\left\lbrace y_{i},p_{yi} W\right\rbrace+\left\lbrace z_{i},p_{zi} W\right\rbrace \right)\\
\fl &&+ D \sum_{i=1}^N m_i \left(\left\lbrace x_i,  \left\lbrace x_i,W\right\rbrace \right\rbrace+\left\lbrace y_i,  \left\lbrace y_i,W\right\rbrace \right\rbrace+\left\lbrace z_i,  \left\lbrace z_i,W\right\rbrace \right\rbrace\right),
  \label{poisson}
\end{eqnarray}
where $\vec{r}_i=(x_i,y_i,z_i)$, $\vec{p}_i=(p_{xi},p_{yi},p_{zi})$, and we defined the Hamiltonian function as $\mathcal{H}=K_0 + V=K_0 +V_0-\sum_{i=1}^N\vec{f}_i\cdot \vec{r}_i=\mathcal{H}_0-\sum_{i=1}^N\vec{f}_i\cdot \vec{r}_i$, where $\mathcal{H}_0=K_0 +V_0$. Here, the three terms of drift (Liouville), friction and noise can be easily recognized.  

The asymptotic behavior of Eq. (\ref{poisson}) for large times is characterized by the canonical or Gibbs distribution \cite{gibbs}. Indeed, if the forces $\vec{f}_i$ are constant in time and the integral defining the classical partition function
\begin{eqnarray}\fl
 \label{trenta}
 Z_{cl} = \int_{\mathcal{A}} \int_{\mathbb{R}^{3N}} e^{- \frac{\beta}{D}\mathcal{H}(\vec{q},\vec{p})} d\vec{q} d\vec{p}
\end{eqnarray}
is convergent (with 
$\vec{q}=(\vec{r}_1,...,\vec{r}_N) \in \mathcal{A} \subset \mathbb{R}^{3N}$  and $\vec{p}=(\vec{p}_1,...,\vec{p}_N) \in \mathbb{R}^{3N}$), then the asymptotic solution of Eq. (\ref{poisson}) is given by the Gibbs distribution in 
phase space
\begin{eqnarray}\fl
 \label{trentuno}
 W_{eq}(\vec{q},\vec{p}) = \frac{1}{Z_{cl}} e^{- \frac{\beta}{D}\mathcal{H}(\vec{q},\vec{p})}.
\end{eqnarray}
This can be easily proved by substitution. 
This asymptotic solution allows the identification of the diffusion constant through the expression $D =  k_B T \beta$, referred to as classical Einstein fluctuation-dissipation relation \cite{coffey,risken}.

We define the internal energy $\mathcal{E}$ of the system as the average value (with respect of the probability density defined by Eq.(\ref{poisson})) of the sum of kinetic energy and potential energy $\mathcal{E}=\mathbb{E} \left\lbrace K_0+V_0\right\rbrace$ and we calculate the rate $\frac{d\mathcal{E}}{dt}$ as follows
\begin{eqnarray}\fl
\frac{d\mathcal{E}}{dt}=\sum_{i=1}^N\vec{f}_i\cdot\mathbb{E}\left\lbrace \vec{v}_i\right\rbrace+2\beta\left(\frac{3}{2}
N k_B T -\mathbb{E}\left\lbrace K_0\right\rbrace\right)=\frac{d\mathbb{E}\left\lbrace L \right\rbrace}{dt}+
\frac{d\mathbb{E}\left\lbrace Q \right\rbrace}{dt}.
\label{1pr}
\end{eqnarray}
This expression represents the first law of thermodynamics,
from which we can identify the rate of average work $\frac{d\mathbb{E}\left\lbrace L \right\rbrace}{dt}$ done on the system with the average power $\sum_{i=1}^N\vec{f}_i\cdot\mathbb{E}\left\lbrace \vec{v}_i\right\rbrace$, and the remaining term with the rate of average heat $\frac{d\mathbb{E}\left\lbrace Q \right\rbrace}{dt}$ entering the system.
This identification is consistent with the Sekimoto definition of heat for a stochastic trajectory \cite{seki1,seki2}.
We observe that the heat flux is zero when the classical equipartition of energy is satisfied. At equilibrium in fact each of the $3N$ quadratic terms of the kinetic energy takes on the value $\frac{k_BT}{2}$.

In order to substantiate the previous explicit expressions of the heat rate, we can obtain the second law of thermodynamics by introducing the Gibbs entropy of the system as
\begin{eqnarray}\fl
\mathcal{S}=-k_B \mathbb{E}\left\lbrace \log W \right\rbrace =-k_B\int_{\mathbb{R}^{3N}}\int_{\mathcal{A}}W\log W d\vec{q} d\vec{p}.
\label{entr}
\end{eqnarray}
This expression means that the microscopic (non-averaged) entropy along a given system trajectory is defined as $-k_B  \log W $, consistently with Refs.\cite{sei1,sei2,sei3}.
The evolution equation can be rewritten as
\begin{eqnarray}\fl
 &&\frac{\partial W}{\partial t} = \left\lbrace \mathcal{H},W \right\rbrace  -\sum_{i=1}^N\frac{\partial \vec{J}_i}{\partial \vec{p}_i},
\end{eqnarray}
where $\vec{J}_i=-\beta W \vec{p}_i-k_BT\beta m_i\frac{\partial W}{\partial \vec{p}_i}$. We note that the derivative $\frac{\partial \vec{J}_i}{\partial \vec{p}_i}$ represents a divergence with respect to the components of $\vec{p}_i$.
These premises lead to the entropy rate in the form
\begin{eqnarray}\fl
\frac{d\mathcal{S}}{dt}&=&\frac{1}{T} \frac{d\mathbb{E}\left\lbrace Q\right\rbrace }{dt}+\frac{1}{\beta T} \int_{\mathbb{R}^{3N}}\int_{\mathcal{A}}\frac{1}{W}\sum_{i=1}^N\frac{\vec{J}_i\cdot\vec{J}_i}{m_i } d\vec{q} d\vec{p}.
\label{2pr}
\end{eqnarray}
We observe that the second term (entropy production) is always non-negative since it is constituted by a quadratic expression \cite{sch,espo1,espo2,tome0,tome1,tome2,giorda}.
Therefore, we obtain the second law of the thermodynamics in the classical form  \begin{eqnarray}\fl
\frac{d\mathcal{S}}{dt}& \geq &\frac{1}{T} \frac{d\mathbb{E}\left\lbrace Q\right\rbrace }{dt},
\label{2prfin}
\end{eqnarray}
where the equality is satisfied only for quasi-static transformations, evolving not far from the thermodynamic equilibrium (for further details, see Ref.\cite{giorda}). 
It is interesting to note that this thermodynamic structure is preserved even when we treat a holonomic system with arbitrary mechanical constraints \cite{annphys,giorda}. 
This scheme can be generalized to introduce the overdamped approximation \cite{annphys,giorda,pan}, and the case with multiple reservoirs \cite{mura}.

In the continuation of the article we will develop a similar procedure for a quantum system in contact with a thermal bath. We begin by studying thermal noise at the quantum level in the next section.

\section{Thermal noise in quantum mechanics}

\subsection{A stochastic Hamiltonian}

Let us now consider a quantum system that is subjected only to the action of stochastic forces. The noise forces $\vec{f}_{si}=\sqrt{D m_i} \vec n_i(t)$, included in Eq. (\ref{totforce}), can be associated with a potential energy $V_{si}=-\sqrt{D m_i} \vec n_i(t)\cdot\vec{r}_i$, such that $\vec{f}_{si}=-\frac{\partial V_{si}}{\partial \vec{r}_i}$ (here, the subscript $s$ means stochastic, and the subscript $i$ indicates the particle number). 
We assume that there are no external forces applied ($\vec{f}_i=0,\,\forall i$) and no dissipative mechanisms present, and
introduce the Hamiltonian $\mathcal{H}_0=K_0 + V_0$. We then study the system described by the stochastic Hamiltonian $\mathcal{H}_s=\mathcal{H}_0-\sum_{i=1}^N\sqrt{D m_i} \vec n_i(t)\cdot\vec{r}_i$.
We introduce the stochastic vector $\vec{n}=(\vec{n}_1,...,\vec{n}_N)\in\mathbb{R}^{3N}$, and the quantities $A_1=-\sqrt{D m_1}x_1$, $A_2=-\sqrt{D m_1}y_1$, $A_3=-\sqrt{D m_1}z_1$, $A_4=-\sqrt{D m_2}x_2$,..., $A_{3N}=-\sqrt{D m_N}z_N$.
Therefore, the stochastic Hamiltonian can be written as 
\begin{equation}\fl
    \mathcal{H}_s=\mathcal{H}_0+\sum_{k=1}^{3N}A_kn_k.
    \label{hamstoc}
\end{equation}
We consider now the Schrödinger equation describing 
this stochastic system
\begin{equation}\fl
    \mathcal{H}_s\Psi=i\hbar\frac{\partial \Psi}{\partial t},
    \label{schro}
\end{equation}
and we introduce an orthonormal basis $\left\lbrace \varphi_n(\vec{q}):\mathbb{R}^{3N}\to\mathbb{C} \right\rbrace$ in the Hilbert space of the wave-functions $\Psi(\vec{q},t)$, equipped with scalar product denoted by the symbol $\langle \cdot\vert\cdot \rangle$. 
We have that $\langle \varphi_n\vert\varphi_m \rangle=\int\varphi_n^*\varphi_md\vec{q}=\delta_{nm} $, and an arbitrary wave-function can be 
expanded as $\Psi(\vec{q},t)=a_n(t)\varphi_n(\vec{q})$ (we adopt the Einstein summation notation for the elements of the basis), with coordinates $a_n=\langle \varphi_n\vert\Psi \rangle\in\mathbb{C}$.
By substituting $\Psi=a_n\varphi_n$ in Eq. (\ref{schro}), and performing the scalar product by $\varphi_m$, we get
\begin{equation}\fl
    \langle \varphi_m\vert\mathcal{H}_s\varphi_n \rangle a_n=i\hbar\frac{da_m}{dt},
\end{equation}
or, equivalently,
\begin{equation}\fl
    \frac{da_m}{dt}=\frac{1}{i\hbar}\langle \varphi_m\vert\mathcal{H}_0\varphi_n \rangle a_n+\frac{1}{i\hbar}\sum_{k=1}^{3N}\langle \varphi_m\vert A_k\varphi_n \rangle a_n n_k,
    \label{stochschro}
\end{equation}
where the operators $A_k$ have been previously defined. It is clear from this equation that in this case the noises act {\it multiplicatively} on the system.
In  \ref{appa}, we prove the following general property: given the stochastic differential equation 
\begin{equation}\fl
    \frac{d\vec{y}}{dt}=\left(C+\sum_{j=1}^\mathfrak{m}D_jn_j(t)\right)\vec{y},
    \label{geomulti}
\end{equation}
we can determine the expectation value $\mathbb{E}\left\lbrace\vec{y}\right\rbrace$ of the vector $\vec{y}\in\mathbb{C}^\mathfrak{n}$ with the ordinary differential equation
\begin{equation}\fl
    \frac{d\mathbb{E}\left\lbrace\vec{y}\right\rbrace}{dt}=\left(C+2\alpha\sum_{j=1}^\mathfrak{m}D_j^2\right)\mathbb{E}\left\lbrace\vec{y}\right\rbrace,
    \label{geomultiave}
\end{equation}
where $\alpha$ is the discretization parameter introduced in Eq. (\ref{fokpla}).
Here, $C$ and $D_j$ are arbitrary complex matrices $\mathfrak{n}\times\mathfrak{n}$, and the real noises $n_j(t)$ satisfy the properties described above.
This result generalizes Eq. (13) of Ref.\cite{geom} for the scalar geometric Brownian motion to the case of the multidimensional geometric Brownian motion (and complex variables).
If we identify the vector $\vec{y}$ with the vector of the coordinates $a_m$ in Eq. (\ref{stochschro}), this property eventually leads to
\begin{eqnarray}\fl
    \frac{d\mathbb{E}\left\lbrace a_m\right\rbrace}{dt}=\frac{1}{i\hbar}\langle \varphi_m\vert\mathcal{H}_0\varphi_n \rangle \mathbb{E}\left\lbrace a_n\right\rbrace -\frac{2\alpha}{\hbar^2}\sum_{k=1}^{3N}\langle \varphi_m\vert A_k\varphi_s \rangle \langle \varphi_s\vert A_k\varphi_n \rangle \mathbb{E}\left\lbrace a_n\right\rbrace.
    \label{stochschroave}
\end{eqnarray}
It means that the expectation value (with respect to the thermal noise) of the wave-function should satisfy the equation
\begin{equation}\fl
    \frac{\partial \mathbb{E}\left\lbrace \Psi\right\rbrace}{\partial t}=\frac{1}{i\hbar}\mathcal{H}_0\mathbb{E}\left\lbrace \Psi\right\rbrace-\frac{2\alpha}{\hbar^2}\sum_{k=1}^{3N}A_k^2\mathbb{E}\left\lbrace \Psi\right\rbrace,
\end{equation}
where $\mathcal{H}_0$, $A_k$ (and then $A_k^2$) are Hermitian operators.
The evolution of the wave-function norm is described by
\begin{eqnarray}\fl
\nonumber
\frac{d}{dt}\langle \mathbb{E}\left\lbrace \Psi\right\rbrace\vert\mathbb{E}\left\lbrace \Psi\right\rbrace \rangle&=&\left\langle \left.\frac{\partial \mathbb{E}\left\lbrace \Psi\right\rbrace}{\partial t}\right\vert\mathbb{E}\left\lbrace \Psi\right\rbrace \right\rangle+\left\langle \mathbb{E}\left\lbrace \Psi\right\rbrace\left\vert\frac{\partial \mathbb{E}\left\lbrace \Psi\right\rbrace}{\partial t}\right.\right\rangle\\
\nonumber
&=&\left\langle \left.\frac{1}{i\hbar}\mathcal{H}_0\mathbb{E}\left\lbrace \Psi\right\rbrace-\frac{2\alpha}{\hbar^2}\sum_{k=1}^{3N}A_k^2\mathbb{E}\left\lbrace \Psi\right\rbrace\right\vert\mathbb{E}\left\lbrace \Psi\right\rbrace \right\rangle\\
&&+\left\langle \mathbb{E}\left\lbrace \Psi\right\rbrace\left\vert\frac{1}{i\hbar}\mathcal{H}_0\mathbb{E}\left\lbrace \Psi\right\rbrace-\frac{2\alpha}{\hbar^2}\sum_{k=1}^{3N}A_k^2\mathbb{E}\left\lbrace \Psi\right\rbrace\right.\right\rangle.
\end{eqnarray}
By using the Hermitian character of $\mathcal{H}_0$ and $A_k^2$, we get
\begin{eqnarray}
\nonumber
\fl\frac{d}{dt}\langle \mathbb{E}\left\lbrace \Psi\right\rbrace\vert\mathbb{E}\left\lbrace \Psi\right\rbrace \rangle&=&-\frac{4\alpha}{\hbar^2}\sum_{k=1}^{3N}\left\langle \mathbb{E}\left\lbrace \Psi\right\rbrace\left\vert A_k^2\mathbb{E}\left\lbrace \Psi\right\rbrace\right.\right\rangle=-\frac{4\alpha}{\hbar^2}\sum_{k=1}^{3N}\left\langle A_k\mathbb{E}\left\lbrace \Psi\right\rbrace\left\vert A_k\mathbb{E}\left\lbrace \Psi\right\rbrace\right.\right\rangle\\
\fl&=&-\frac{4\alpha}{\hbar^2}\sum_{k=1}^{3N}\Vert A_k\mathbb{E}\left\lbrace \Psi\right\rbrace\Vert^2\le 0.
\end{eqnarray}
The norm of the averaged wave-function is therefore not conserved during the system evolution (except for the It\^{o} case with $\alpha=0$, which preserves the norm trivially as it eliminates the effects of noise).
This result is important since it explains that we cannot use the Schrödinger equation with stochastic Hamiltonians
to achieve our aim, 
because the state of the system cannot remain pure. 
This means that we must describe the evolution of a mixed state and this must be done through the density operator and the corresponding Liouville-von Neumann equation \cite{landauq,cohen}.

\subsection{Dynamics of the density operator}

A mixed state is described by $M$ wave-functions $\Psi_1,...,\Psi_M$, associated with the corresponding probabilities $p_1,...,p_M$, with $\sum_{j=1}^Mp_j=1$.
The density operator is therefore defined as $\rho(\vec{q},\vec{q}',t)=\sum_{j=1}^Mp_j\Psi_j(\vec{q},t)\Psi_j^*(\vec{q}',t)$. 
The 
expectation value of an observable $f$ is then calculated as $\mathbb{E}\left\lbrace f\right\rbrace=\sum_{j=1}^Mp_j\langle \Psi_j\vert f \Psi_j \rangle=\int f\rho\vert_{\vec{q}'=\vec{q}}d\vec{q}$, where $f$ represents the operator acting only on the variables $\vec{q}$.
If we adopt the orthonormal basis $\left\lbrace \varphi_n(\vec{q}):\mathbb{R}^{3N}\to\mathbb{C} \right\rbrace$, we have that $\Psi_j=a_{kj}\varphi_k$, with $a_{kj}=\langle \varphi_k\vert  \Psi_j \rangle$.
Hence, the expectation value of $f$ can be written as $\mathbb{E}\left\lbrace f\right\rbrace=\int \sum_{j=1}^Mp_ja_{kj}f\varphi_ka_{hj}^*\varphi_h^*d\vec{q}=\rho_{kh}f_{hk}=\mbox{Tr}(\rho f)$, where we identified the representations $\rho_{kh}=\sum_{j=1}^Mp_ja_{kj}a_{hj}^*$, and $f_{hk}=\int \varphi_h^*f\varphi_kd\vec{q}=\langle \varphi_h\vert f \varphi_k \rangle$.

The density matrix $\rho_{kh}$ satisfies certain properties that will have to be fulfilled during the time evolution \cite{landauq,cohen}: 
\

\begin{enumerate}[(i)]

\item  its trace is unitary, indeed $\mbox{Tr}\rho=\rho_{kk}=\sum_{j=1}^Mp_ja_{kj}a_{kj}^*=\sum_{j=1}^Mp_j\langle \Psi_j\vert  \Psi_j \rangle=1$; 
\\

\item the diagonal elements 
are non-negative since we can write $\rho_{kk}=\sum_{j=1}^Mp_ja_{kj}a_{kj}^*=\sum_{j=1}^Mp_j\vert\langle \varphi_k\vert  \Psi_j \rangle\vert^2\ge 0$ (without the sum over $k$); 
\\

\item the density matrix is Hermitian, in fact we have that $\rho_{kh}=\sum_{j=1}^Mp_ja_{kj}a_{hj}^*$, and then $\rho_{hk}^*=\sum_{j=1}^Mp_ja_{hj}^*a_{kj}=\rho_{kh}$, or $\rho^{T*}=\rho$ ($T$ means ``transposed''); 
\\

\item the density matrix is positive-definite since $v^{T*}\rho v=v_k^*\rho_{kh}v_h=\sum_{j=1}^Mp_jv_k^*a_{kj}v_ha_{hj}^*=\sum_{j=1}^Mp_j(v_ka_{kj}^*)^*(v_ha_{hj}^*)=\sum_{j=1}^Mp_j\vert v_ka_{kj}^*\vert^2>0$.
\

\end{enumerate}

For a system defined by the Hamiltonian $\mathcal{H}_s$ the time evolution of the density matrix is described by the Liouville-von Neumann equation  \cite{landauq,cohen}
\begin{equation}\fl
    \frac{d\rho}{dt}=\frac{1}{i\hbar}\left[\mathcal{H}_s,\rho\right],
\end{equation}
where $\left[A,B\right]=AB-BA$ is the commutator of $A$ and $B$.
This equation describes the evolution of a mixed state, taking into consideration the Schrödinger equation for each state $\Psi_j$.
If we consider the stochastic Hamiltonian in Eq. (\ref{hamstoc}), we get
\begin{equation}\fl
    \frac{d\rho}{dt}=\frac{1}{i\hbar}\left[\mathcal{H}_0,\rho\right]+\frac{1}{i\hbar}\sum_{k=1}^{3N}\left[A_k,\rho\right]n_k.
    \label{stochrho}
\end{equation}
This approach is somewhat analogous to that developed in Ref.\cite{kiely2021}.
This stochastic equation for the density matrix has a form similar to Eq. (\ref{geomulti}), but it describes the dynamics of a matrix and not of a vector. Therefore, we need to introduce a transformation to rewrite it with an unknown vector.  
We define first the Kronecker product of two matrices ${A}$ and ${B}$ through the block matrix
\begin{eqnarray}\fl
{A}\otimes {B}=\left[\begin{array}{cccc}
a_{11}{B} &a_{12}{B} &a_{13}{B} &\cdot\cdot\cdot\\
a_{21}{B} &a_{22}{B} &a_{23}{B} &\cdot\cdot\cdot\\
a_{31}{B} &a_{32}{B} &a_{33}{B} & \cdot\cdot\cdot\\
\vdots&\vdots&\vdots&\ddots\\
\end{array} \right].
\end{eqnarray} 
This operation is non-commutative and is useful to convert equations like Eq. (\ref{stochrho}) to the standard vector representation. 
To do this, we also need to define the vectorization of a matrix. This operation  converts a matrix ${A}$ into a column vector $\hat{{A}}$ by juxtaposing the consecutive rows of the matrix and transposing the result
\begin{eqnarray}
\fl{A}=\left[\begin{array}{cccc}
a_{11} &a_{12} &a_{13} &\cdot\cdot\cdot\\
a_{21} &a_{22} &a_{23} &\cdot\cdot\cdot\\
a_{31} &a_{32} &a_{33} &\cdot\cdot\cdot\\
\vdots&\vdots&\vdots&\ddots\\
\end{array} \right] \Rightarrow \hat{{A}}=\left[a_{11}\, a_{12}\, a_{13}\, ... \,a_{21}\, a_{22}\, a_{23}\, ... \,a_{31}\, a_{32}\, a_{33}\, ...\right]^T.
\end{eqnarray}
The important relation between the Kronecker product and vectorization is given by the following properties
\begin{eqnarray}
\label{kron1}
\fl{A}&=&{B}{C}\Rightarrow\hat{{A}}=\left({B}\otimes{I}\right) \hat{{C}}=\left({I}\otimes{C}^T\right) \hat{{B}},\\
\label{kron2}
\fl{Z}&=&{A}{B}{C}\Rightarrow \hat{{Z}}=\left({A}\otimes{I}\right) \left({I}\otimes{C}^T\right)\hat{{B}},
\end{eqnarray} 
where ${I}$ is the identity matrix.
These properties allow us to state that the vectorization $\hat{\rho}$ of $\rho$ is described by the following equation
\begin{eqnarray}
\fl
\frac{d\hat{\rho}}{dt}=\frac{1}{i\hbar}\left(\mathcal{H}_0\otimes I-I\otimes \mathcal{H}_0^T\right)\hat{\rho}+\frac{1}{i\hbar}\sum_{k=1}^{3N}\left(A_k\otimes I-I\otimes A_k^T\right)n_k\hat{\rho},
\end{eqnarray}
which is exactly of the form given in Eq. (\ref{geomulti}).
Therefore, we can obtain the evolution of the 
expectation value $\mathbb{E}\left\lbrace\hat{\rho}\right\rbrace$ through Eq. (\ref{geomultiave}).
This yields the following result
\begin{eqnarray}\fl
\frac{d\mathbb{E}\left\lbrace\hat{\rho}\right\rbrace}{dt}=\frac{1}{i\hbar}\left(\mathcal{H}_0\otimes I-I \otimes\mathcal{H}_0^T\right)\mathbb{E}\left\lbrace\hat{\rho}\right\rbrace
-\frac{2\alpha}{\hbar^2}\sum_{k=1}^{3N}\left(A_k\otimes I-I\otimes A_k^T\right)^2\mathbb{E}\left\lbrace\hat{\rho}\right\rbrace.
\end{eqnarray}
Using again the properties of the Kronecker product and the vectorization process, we can rewrite the equation in the matrix formalism as follows
\begin{equation}\fl
    \frac{d\mathbb{E}\left\lbrace{\rho}\right\rbrace}{dt}=\frac{1}{i\hbar}\left[\mathcal{H}_0,\mathbb{E}\left\lbrace{\rho}\right\rbrace\right]-\frac{2\alpha}{\hbar^2}\sum_{k=1}^{3N}\left[A_k,\left[A_k,\mathbb{E}\left\lbrace{\rho}\right\rbrace\right]\right].
    \label{stochrhofinak}
\end{equation}
We can also remember the definition of the  operators $A_k$, and thus write the explicit form of this equation
\begin{eqnarray}\nonumber\fl
    \frac{d\mathbb{E}\left\lbrace{\rho}\right\rbrace}{dt}=\frac{1}{i\hbar}\left[\mathcal{H}_0,\mathbb{E}\left\lbrace{\rho}\right\rbrace\right]-\frac{2\alpha D}{\hbar^2}\sum_{k=1}^{N}m_k\left(\vphantom{\frac{1}{2}}\left[x_k,\left[x_k,\mathbb{E}\left\lbrace{\rho}\right\rbrace\right]\right]
    +\left[y_k,\left[y_k,\mathbb{E}\left\lbrace{\rho}\right\rbrace\right]\right]+\left[z_k,\left[z_k,\mathbb{E}\left\lbrace{\rho}\right\rbrace\right]\right]\vphantom{\frac{1}{2}}\right).\\
    \fl\label{stochrhofin}
\end{eqnarray}
We emphasize that the quantum terms concerning the thermal noise are closely analogous to those obtained in classical mechanics, see the third line of Eq. (\ref{poisson}). 
Indeed, performing the formal substitution  $\left\{\cdot ,\cdot \right\}\rightarrow \frac {1}{i\hbar}\left[\cdot ,\cdot \right]$ (Poisson brackets $\to$ commutators: canonical quantization), the noise terms become like those in Eq. (\ref{stochrhofin}).
The presence of the constant $2\alpha$ will be discussed below, and is concerned with the stochastic interpretation adopted.
In conclusion, Eq. (\ref{stochrhofin}) describes the evolution of a mixed state for a quantum system with stochastic terms.  
We want to emphasize that this equation was obtained rigorously and that the analogy with the classical case was only observed a posteriori. 
It is important to note that the expectation value symbol used in $\mathbb{E} \left\lbrace \rho\right\rbrace$ refers to the average with respect to thermal noise, and should not be confused with the quantum average of an observable, which now becomes $\mathbb{E}\left\lbrace f\right\rbrace=\mbox{Tr}(\mathbb{E} \left\lbrace \rho\right\rbrace f)$. In $\mathbb{E}\left\lbrace f\right\rbrace$, the expectation value symbol refers of course to both quantum indeterminacy and thermal noise.

To achieve proper thermodynamic behavior, we must add a dissipation mechanism to the thermal fluctuations just introduced.
In fact, the energy of the system $\mathcal{E}=\mbox{Tr}(\mathcal{H}_0\mathbb{E} \left\lbrace \rho\right\rbrace)$ with only the noise terms would always be increasing as seen from 
the following direct evaluation. In a first step, we have
\begin{eqnarray}
\nonumber
\fl\frac{d\mathcal{E}}{dt}&=&\frac{d\mbox{Tr}(\mathcal{H}_0\mathbb{E} \left\lbrace \rho\right\rbrace)}{dt}=\mbox{Tr}\left(\mathcal{H}_0\frac{d\mathbb{E}\left\lbrace{\rho}\right\rbrace}{dt}\right)=-\frac{2\alpha D}{\hbar^2}\sum_{k=1}^{N}m_k\mbox{Tr}\left(\mathcal{H}_0\left[x_k,\left[x_k,\mathbb{E}\left\lbrace{\rho}\right\rbrace\right]\right]\right)+...\\
\fl&=&-\frac{2\alpha D}{\hbar^2}\sum_{k=1}^{N}m_k\mbox{Tr}\left(\left[\mathcal{H}_0,x_k\right]\left[x_k,\mathbb{E}\left\lbrace{\rho}\right\rbrace\right]\right)+...,
\end{eqnarray}
where the ellipsis represents the $y$ and $z$ terms. Here, we have used the definition of the commutator and the cyclic property of the trace operation. 
In a second step, we now remember that $\left[\mathcal{H}_0,x_k\right]=-i\hbar p_{xk}/m_k$ (see below for details), and 
get
\begin{eqnarray}
\nonumber
\fl\frac{d\mathcal{E}}{dt}&=&i\frac{2\alpha D}{\hbar}\sum_{k=1}^{N}\mbox{Tr}\left(p_{xk}\left[x_k,\mathbb{E}\left\lbrace{\rho}\right\rbrace\right]\right)+...=i\frac{2\alpha D}{\hbar}\sum_{k=1}^{N}\mbox{Tr}\left(\left[p_{xk},x_k\right]\mathbb{E}\left\lbrace{\rho}\right\rbrace\right)+...
\\
\fl&=&{2\alpha D}\sum_{k=1}^{N}\mbox{Tr}\left(\mathbb{E}\left\lbrace{\rho}\right\rbrace\right)+...=6N\alpha D,
\end{eqnarray}
where we used the canonical commutator
$\left[p_{xk},x_k\right]=-i\hbar$.
This result shows that the energy increases linearly over time if we include thermal fluctuations alone. To maintain energy finite we must then add dissipation terms, equivalent of Langevin's classic thermal bath.
This will be described in the next section.

\section{Dissipation in quantum mechanics}
\label{dissi}

In order to introduce a dissipation 
mechanism -- friction -- into the previously obtained equation, we exploit the analogy with the classical approach.
Following this principle, we  develop the quantum counterpart of the second line of Eq. (\ref{poisson}), by  transforming the terms $p_{xk}W$, $p_{yk}W$, and $p_{zk}W$ into 
Hermitian operators. 
Since the product of two Hermitian operators is not necessarily Hermitian, but their symmetrization is always Hermitian,
we substitute them with the quantum symmetrizations  $\frac{1}{2}(\Theta_{xk}\mathbb{E}\left\lbrace{\rho}\right\rbrace+\mathbb{E}\left\lbrace{\rho}\right\rbrace\Theta_{xk})$, $\frac{1}{2}(\Theta_{yk}\mathbb{E}\left\lbrace{\rho}\right\rbrace+\mathbb{E}\left\lbrace{\rho}\right\rbrace\Theta_{yk})$, and $\frac{1}{2}(\Theta_{zk}\mathbb{E}\left\lbrace{\rho}\right\rbrace+\mathbb{E}\left\lbrace{\rho}\right\rbrace\Theta_{zk})$, where $\Theta_{xk}$, $\Theta_{yk}$, and $\Theta_{zk}$ are Hermitian operators to be determined ($\forall k=1..N$), taking the role of classical momenta. 
By adding the friction terms to Eq. (\ref{stochrhofin}), we obtain the complete equation
\begin{eqnarray}
\nonumber
  \fl  \frac{d\mathbb{E}\left\lbrace{\rho}\right\rbrace}{dt}&=&\frac{1}{i\hbar}\left[\mathcal{H}_0,\mathbb{E}\left\lbrace{\rho}\right\rbrace\right]-\frac{2\alpha D}{\hbar^2}\sum_{k=1}^{N}m_k\left(\vphantom{\frac{1}{2}}\left[x_k,\left[x_k,\mathbb{E}\left\lbrace{\rho}\right\rbrace\right]\right]+\left[y_k,\left[y_k,\mathbb{E}\left\lbrace{\rho}\right\rbrace\right]\right]+\left[z_k,\left[z_k,\mathbb{E}\left\lbrace{\rho}\right\rbrace\right]\right]\vphantom{\frac{1}{2}}\right)\\
  \nonumber
    \fl&&+\frac{\beta}{2i\hbar}\sum_{k=1}^{N}\left(\vphantom{\frac{1}{2}}\left[x_k,\Theta_{xk}\mathbb{E}\left\lbrace{\rho}\right\rbrace+\mathbb{E}\left\lbrace{\rho}\right\rbrace\Theta_{xk}\right]+\left[y_k,\Theta_{yk}\mathbb{E}\left\lbrace{\rho}\right\rbrace+\mathbb{E}\left\lbrace{\rho}\right\rbrace\Theta_{yk}\right]\right.\\
    \fl&&\left.+\left[z_k,\Theta_{zk}\mathbb{E}\left\lbrace{\rho}\right\rbrace+\mathbb{E}\left\lbrace{\rho}\right\rbrace\Theta_{zk}\right]\vphantom{\frac{1}{2}}\right),
    \label{main}
\end{eqnarray}
where we now have to study the structure of the friction operators $\Theta_{xk}$, $\Theta_{yk}$, and $\Theta_{zk}$, $\forall k=1..N$, and the fluctuation-dissipation relation linking the diffusion constant $D$ with the friction coefficient $\beta$.
For the following developments, it is useful to rewrite the previous equation in the more compact form
\begin{eqnarray}\fl
\nonumber
    \frac{d\mathbb{E}\left\lbrace{\rho}\right\rbrace}{dt}&=&\frac{1}{i\hbar}\left[\mathcal{H}_0,\mathbb{E}\left\lbrace{\rho}\right\rbrace\right]-\frac{2\alpha D}{\hbar^2}\sum_{k=1}^{N}m_k\left(\sum_{s=x,y,z}\left[r_{sk},\left[r_{sk},\mathbb{E}\left\lbrace{\rho}\right\rbrace\right]\right]\right)\\
     \label{mainbis}\fl
    &&+\frac{\beta}{2i\hbar}\sum_{k=1}^{N}\left(\sum_{s=x,y,z}\left[r_{sk},\Theta_{sk}\mathbb{E}\left\lbrace{\rho}\right\rbrace+\mathbb{E}\left\lbrace{\rho}\right\rbrace\Theta_{sk}\right]\right),
\end{eqnarray}
where we defined $\vec{r}_k=(x_k,y_k,z_k)=(r_{xk},r_{yk},r_{zk})$.
We have now to find the Hermitian friction operators $\Theta_{sk}$, $s=x,y,z$, $k=1,...,N$, in such a way that the asymptotic behavior of the equation is described by the canonical quantum distribution
\begin{equation}\fl
    \lim_{t\to\infty}\mathbb{E}\left\lbrace{\rho}\right\rbrace=\mathbb{E}\left\lbrace{\rho}\right\rbrace_{eq}=\frac{1}{Z_{qu}}e^{-\frac{\mathcal{H}_0}{k_BT}},
\end{equation}
where $Z_{qu}$ is the quantum partition function
\begin{equation}\fl
 Z_{qu}=\mbox{Tr}\left(e^{-\frac{\mathcal{H}_0}{k_BT}}\right). 
\end{equation}
We will see how this assumption leads to friction operators similar to momentum operators. An interesting alternative solution has been proposed in the literature and is based on the substitution $p_{sk}W\to\frac{1}{2}(g_{sk}^\dagger\mathbb{E}\left\lbrace{\rho}\right\rbrace+\mathbb{E}\left\lbrace{\rho}\right\rbrace g_{sk})$, where non-Hermitian friction operators $g_{sk}$ are considered (the symbol $\dagger$ means ``adjoint operator'') \cite{oliveira2016,oliveira2023,oliveira2024}. This choice leads to some simplifications in the calculations but one cannot associate friction operators with a  {true} physical observable.

In our case, we impose the asymptotic quantum canonical distribution to Eq. (\ref{mainbis}), and we obtain the relation
\begin{eqnarray}\fl
    i\frac{4\alpha Dm_k}{\beta\hbar}\left[r_{sk},\mathbb{E}\left\lbrace{\rho}\right\rbrace_{eq}\right]=\Theta_{sk}\mathbb{E}\left\lbrace{\rho}\right\rbrace_{eq}+\mathbb{E}\left\lbrace{\rho}\right\rbrace_{eq}\Theta_{sk},
    \label{mainequationa}
\end{eqnarray}
or, equivalently,
\begin{eqnarray}\fl
    i\frac{4\alpha Dm_k}{\beta\hbar}\left[r_{sk},e^{-\frac{\mathcal{H}_0}{k_BT}}\right]=\Theta_{sk}e^{-\frac{\mathcal{H}_0}{k_BT}}+e^{-\frac{\mathcal{H}_0}{k_BT}}\Theta_{sk},
    \label{mainequationb}
\end{eqnarray}
for $s=x,y,z$, and $k=1,...,N$.
From a mathematical point of view, this equation in $\Theta_{sk}$ is a matrix equation of the form ${A}{X}+{X}{A}={C}$, which is sometimes called Sylvester or Lyapunov equation, see e.g. Refs.\cite{gant,lanc,brogan}. 
We prove in  \ref{appb} that this equation has the unique solution
\begin{eqnarray}\fl
{X}=-\int_0^{+\infty}e^{{A}\xi}{C}e^{{A}\xi}d\xi,
\end{eqnarray}
if ${A}$ has all eigenvalues with negative real part. 
For solving our equation, we let $X=\Theta_{sk}$, $A=-e^{-\frac{\mathcal{H}_0}{k_BT}}$, and $C=-  i\frac{4\alpha Dm_k}{\beta\hbar}\left[r_{sk},e^{-\frac{\mathcal{H}_0}{k_BT}}\right]$. 
These definitions ensure that the eigenvalues of $A$ are with strictly negative real part. 
Hence, we can write the explicit solution
of Eq. (\ref{mainequationb}) as
\begin{equation}\fl
    \Theta_{sk}=i\frac{4\alpha Dm_k}{\beta\hbar}\int_0^{+\infty}
e^{-\xi e^{-\frac{\mathcal{H}_0}{k_BT}}}\left[r_{sk},e^{-\frac{\mathcal{H}_0}{k_BT}}\right]e^{-\xi e^{-\frac{\mathcal{H}_0}{k_BT}}}d\xi,
\label{integ1}
\end{equation}
where we find a double exponential matrix. 
Although this expression seems rather complicated, it leads to a particularly interesting result when projected onto the energy basis of the Hamiltonian operator.
If we assume that the spectrum of the system is discrete and non-degenerate, we then have that $\mathcal{H}_0\varphi_n(\vec{q})=E_n\varphi_n(\vec{q})$, with $\langle \varphi_n\vert\varphi_m \rangle=\delta_{nm} $.
In this basis, the operator $e^{-\frac{\mathcal{H}_0}{k_BT}}$ is diagonal with elements $e^{-\frac{E_n}{k_BT}}$. 
To simplify the notation, we introduce the quantity $e_n=e^{-\frac{E_n}{k_BT}}>0$. 
Therefore the central matrix $[r_{sk},e^{-\frac{\mathcal{H}_0}{k_BT}}]$ is composed of the following elements
\begin{equation}\fl
[r_{sk},e^{-\frac{\mathcal{H}_0}{k_BT}}]_{pq}=r_{sk,p\ell}e_{\ell} \delta_{\ell q}-e_p\delta_{pj}r_{sk,jq}=r_{sk,pq}(e_q-e_p).
\end{equation}
The structure of the friction operator in Eq. (\ref{integ1}) assumes the form 
\begin{eqnarray}\fl
\nonumber
    \Theta_{sk,\ell j}&=&i\frac{4\alpha Dm_k}{\beta\hbar}\int_0^{+\infty}
e^{-e_{\ell}\xi}\delta_{\ell p}r_{sk,pq}(e_q-e_p)e^{-e_j\xi}\delta_{qj}d\xi\\
\fl
&=&i\frac{4\alpha Dm_k}{\beta\hbar}\int_0^{+\infty}
e^{-e_{\ell}\xi}r_{sk,\ell j}(e_j-e_{\ell})e^{-e_j\xi}d\xi=i\frac{4\alpha Dm_k}{\beta\hbar}r_{sk,\ell j}\frac{e_j-e_{\ell}}{e_j+e_{\ell}}.
\label{friene}
\end{eqnarray}
To further simplify this expression, we prove a simple relation between the position coefficients $r_{sk,\ell j}=\langle\varphi_{\ell}\vert r_{sk}\varphi_j \rangle$ and the momentum coefficients $p_{sk,\ell j}=-i\hbar\langle\varphi_{\ell}\vert \frac{\partial}{\partial r_{sk}}\varphi_j \rangle$. 
We start by considering the canonical commutator $[r_{sk},p_{sk}]=i\hbar$, and then we get $[r_{sk},p_{sk}^2]=r_{sk}p_{sk}^2-p_{sk}^2r_{sk}+p_{sk}r_{sk}p_{sk}-p_{sk}r_{sk}p_{sk}=[r_{sk},p_{sk}]p_{sk}+p_{sk}[r_{sk},p_{sk}]=2i\hbar p_{sk}$. 
Hence, we can write the commutator $[r_{sk},\mathcal{H}_0]=[r_{sk},K_0+V_0]=\frac{1}{2m_k}[r_{sk},p_{sk}^2]=\frac{i\hbar}{m_k}p_{sk}$. 
Projecting the latter relationship onto the energy basis of the system, we obtain $r_{sk,\ell q}\mathcal{H}_{0,qj}-\mathcal{H}_{0,\ell p}r_{sk,pj}=\frac{i\hbar}{m_k}p_{sk,\ell j}$, or equivalently $r_{sk,\ell q}E_q\delta_{qj}-E_{\ell} \delta_{\ell p}r_{sk,pj}=\frac{i\hbar}{m_k}p_{sk,\ell j}$, leading to the result $r_{sk,\ell j}E_j-E_{\ell} r_{sk,\ell j}=\frac{i\hbar}{m_k}p_{sk,\ell j}$.
Finally, we have proved the direct link between position and momentum $r_{sk,\ell j}=\frac{i\hbar}{m_k}\frac{p_{sk,\ell j}}{E_j-E_{\ell}}$. 
Substituting the latter relationship and the property $\frac{e_j-e_{\ell}}{e_j+e_
{\ell}}=\tanh\left(\frac{E_{\ell}-E_j}{2k_B T}\right)$ into Eq. (\ref{friene}), we get
\begin{eqnarray}\fl
    \Theta_{sk,\ell j}=\frac{2\alpha D}{k_B T\beta}p_{sk,\ell j}\frac{\tanh\left(\frac{E_{\ell}-E_j}{2k_B T}\right)}{\frac{E_{\ell}-E_j}{2k_B T}}.
\label{frienefin}
\end{eqnarray}
To impose the maximal similarity
between $\Theta_{sk}$ and $p_{sk}$ we can assume the fluctuation-dissipation relation
\begin{equation}\fl
    D=\frac{k_B T \beta}{2\alpha}.
    \label{quadiff}
\end{equation}
This means that we can construct the quantum relaxation with any kind of stochastic interpretation, $0<\alpha\le 1$, except the Ito interpretation with $\alpha=0$, which cancels the effects of noise in the multidimensional geometric Brownian process, as mentioned before. 
As a special case, we observe that with the Stratonovich interpretation, $\alpha=1/2$, we restore the Einstein fluctuation-dissipation relation $D =  k_B T \beta$, discussed in the classical formalism.
Once Eq. (\ref{quadiff}) is assumed, the friction operators take the final form
\begin{eqnarray}\fl
    \Theta_{sk,\ell j}=p_{sk,\ell j}\frac{\tanh\left(\frac{E_{\ell}-E_j}{2k_B T}\right)}{\frac{E_{\ell}-E_j}{2k_B T}}.
\label{frienefind}
\end{eqnarray}
This result shows that in quantum relaxation viscous friction is not proportional to the momentum of the particles but to a new operator that resembles momentum but depends on the energy levels of the system and the temperature itself. 
As we will see below, this difference is related to the fact that quantum equipartition is different from classical equipartition.
As a final result, once Eq. (\ref{quadiff}) is assumed, we have the complete evolution of the density matrix governed by the equation 
\begin{eqnarray}\fl
\nonumber
    \frac{d\mathbb{E}\left\lbrace{\rho}\right\rbrace}{dt}&=&\frac{1}{i\hbar}\left[\mathcal{H}_0,\mathbb{E}\left\lbrace{\rho}\right\rbrace\right]-\frac{k_B T \beta}{\hbar^2}\sum_{k=1}^{N}m_k\left(\sum_{s=x,y,z}\left[r_{sk},\left[r_{sk},\mathbb{E}\left\lbrace{\rho}\right\rbrace\right]\right]\right)\\
    \label{maindiff}\fl
    &&+\frac{\beta}{2i\hbar}\sum_{k=1}^{N}\left(\sum_{s=x,y,z}\left[r_{sk},\Theta_{sk}\mathbb{E}\left\lbrace{\rho}\right\rbrace+\mathbb{E}\left\lbrace{\rho}\right\rbrace\Theta_{sk}\right]\right),
\end{eqnarray}
where the friction coefficient $\beta$ controls the rate of relaxation toward quantum thermodynamic equilibrium.
Let us again emphasize that Eq. (\ref{maindiff}) is strictly analogous to Eq. (\ref{poisson}) once we adopt the canonical quantization by replacing Poisson brackets with commutators. The quantum novelty is that new friction operators $\Theta_{sk}$ must be introduced.  
Their final form in an arbitrary basis is given by Eq. (\ref{integ1}) combined with Eq. (\ref{quadiff}), and results in
\begin{equation}\fl
    \Theta_{sk}=i\frac{2 m_k k_B T }{\hbar}\int_0^{+\infty}
e^{-\xi e^{-\frac{\mathcal{H}_0}{k_BT}}}[r_{sk},e^{-\frac{\mathcal{H}_0}{k_BT}}]e^{-\xi e^{-\frac{\mathcal{H}_0}{k_BT}}}d\xi.
\label{integ1a}
\end{equation}
Another different integral form for the friction operator, always in an arbitrary basis, is obtained in  \ref{appc}. It reads as
\begin{equation}\fl
    \Theta_{sk}=\frac{2 }{\pi}\int_{-\infty}^{+\infty}
e^{+i\frac{\mathcal{H}_0}{k_BT}\eta}p_{sk}e^{-i\frac{\mathcal{H}_0}{k_BT}\eta}\log\left[\coth\left(\frac{\pi}{2}\vert\eta\vert\right)\right]d\eta.
\label{integ2a}
\end{equation}
This expression shows again that the friction operator is strongly related to the momentum operator, and it yields again Eq. (\ref{frienefind}) when the energy basis is adopted.  
Moreover, the following expansion is also proved in  \ref{appc}
\begin{eqnarray}\fl
    {\Theta}_{sk}=p_{sk}-\frac{1}{12(k_BT)^2}[\mathcal{H}_0,[\mathcal{H}_0,p_{sk}]]+\frac{1}{120(k_BT)^4}[\mathcal{H}_0,[\mathcal{H}_0,[\mathcal{H}_0,[\mathcal{H}_0,p_{sk}]]]]...,
    \label{threeterms}
\end{eqnarray}
where all coefficients of the development are written in terms of Bernoulli numbers. 
We see that when the thermal energy $k_BT$ is high  enough (that is, much larger than the energy-level separation) the system approaches the classical behavior and in fact ${\Theta}_{sk}\to p_{sk}$.
 {Recall that the Caldeira-Leggett model is obtained from ours by assuming that ${\Theta}_{sk}= p_{sk}$ \cite{caldeira1981,caldeira1983,caldeira1983bis}. 
This means, looking at Eq. (\ref{threeterms}), that this model is approximate and holds only for sufficiently high temperatures.}

 {We obtained the representation of the friction operator on the energy basis only for a discrete, nondegenerate spectrum. Nevertheless, Eqs. (\ref{integ1a}), (\ref{integ2a}) and (\ref{threeterms}) are valid in any case since they are purely operator relations. 
In particular, they are also valid in the continuous spectrum case. For example, for free particles the Hamiltonian function coincides with the kinetic energy and thus it is simply obtained from these relations that ${\Theta}_{sk}= p_{sk}$ (in fact $\mathcal{H}_0$ commutes with $p_{sk}$ in this case). 
However, it is important to note that the continuous spectrum is typically observed in non-confined systems for which there is divergence of the partition function, and thus the canonical asymptotic distribution is actually not reached. The case of a degenerate energy spectrum is discussed in the \ref{appd} for completeness.}

 {The physical origin of our model coincides with that of the classical Langevin equation.
This model describes the dynamics of a particle immersed in a thermal bath by decomposing the  forces acting on it into a deterministic and a stochastic component. 
The deterministic force typically includes conservative interactions, and a linear friction term that accounts for the dissipative effect of the surrounding medium. The stochastic component represents the fluctuating influence of the thermal bath, modeled as a random force with zero mean.
A key physical assumption is that the thermal bath consists of a large number of microscopic degrees of freedom in thermal equilibrium, which interact weakly and rapidly with the particle \cite{lindenberg1990}. 
This separation of timescales justifies modeling the random force as a Gaussian white noise, characterized by a delta-correlated autocorrelation function. The assumption of delta-correlation—implying no memory—gives rise to the Markovian hypothesis, whereby the future evolution of the system depends only on its current state, not its past history \cite{risken,coffey}.
This Markovian behavior is valid when the relaxation time of the bath is much shorter than that of the particle. 
As a result, the bath remains effectively unperturbed by the dynamics of the particle, and its influence can be captured statistically via the fluctuation-dissipation theorem, which relates the noise amplitude to the friction coefficient and temperature. 
This framework yields a stochastic differential equation—the Langevin equation—that captures the mesoscopic dynamics of the system while coarse-graining the microscopic details of the bath. 
The emergence of irreversibility in the Langevin description stems from this coarse-graining over the bath degrees of freedom. While the full microscopic dynamics of the combined system (particle + bath) are time-reversible and governed by Hamiltonian or Newtonian mechanics, the elimination of the bath variables introduces an effective asymmetry in time. 
Friction leads to energy dissipation, while the stochastic term models thermal agitation without recovering the detailed information lost to the bath. 
This asymmetry manifests macroscopically as entropy production and defines the arrow of time in nonequilibrium processes, despite the underlying time-reversible laws at the microscopic scale \cite{maddox1985,lebowitz1993,guff2025}.
The quantum Caldeira-Leggett equation is based exactly on the same physical assumptions of the Langevin model \cite{caldeira1981,caldeira1983,caldeira1983bis}, and in fact it can be obtained directly by applying the canonical quantization to the Fokker-Planck equation (keeping the momentum operator as the friction operator). 
The development of quantum models via canonical quantization of the Fokker-Planck equation has recently emerged as a successful approach to circumvent the need for explicitly modeling the thermal bath, as required in the original formulations by Langevin or Caldeira–Leggett \cite{oliveira2016,oliveira2023,oliveira2024}. 
An alternative approach to introduce friction is based on the Green-Kubo formalism, relating the effective friction coefficient to microscopic dynamics via time-correlation functions. Specifically, the friction is expressed as the time integral of the autocorrelation of the fluctuating force exerted by the thermal bath \cite{friction1,friction2}. 
This approach provides a first-principles derivation of dissipative behavior, linking friction to equilibrium fluctuations through linear response theory. This approach is more complex though more refined.
It is important to remark that the Langevin equation automatically converges to the classical canonical distribution whereas the Caldeira-Leggett equation has asymptotic thermodynamic behavior that does not converge to the quantum canonical distribution. For this reason in this work the quantum friction operator does not coincide with the momentum operator. This modification to the original Caldeira-Leggett model allows us to obtain the correct thermodynamic behavior, as we describe below. We will directly show the comparison between Caldeira-Leggett approach and ours to clarify this point.}

Before studying the development of the quantum stochastic thermodynamics, let us observe that Eq. (\ref{maindiff}) has the exact form of a master equation and meets its basic constraints \cite{happer1972,blum1981,bianco}. Indeed, Eq. (\ref{maindiff}) can be written in the form
\begin{equation}\fl
\frac{d\varrho_{ij}}{dt}=\mathcal{L}_{ijrs}\varrho_{rs}, 
\end{equation} 
where $\mathcal{L}_{ijrs}$ is the complete quantum Langevin relaxation superoperator.
The evolution does not change the trace of the density matrix and therefore $\sum_i\mathcal{L}_{iirs}=0$. This relation can be easily verified for our master equation. Moreover, the density matrix of the system must be Hermitian at all times. Consequently, $\mathcal{L}_{ijrs}^*=\mathcal{L}_{jirs}$, and this constraint is also verified by our equation.
The following two constraints concern the positive definiteness of the density matrix.
The diagonal matrix elements must be non-negative and therefore $\mathcal{L}_{iiii}\le 0$. 
If we  suppose that $\varrho_{ii}=1$ and all other components are zero, then the equation $\frac{d\varrho_{ij}}{dt}=\mathcal{L}_{ijrs}\varrho_{rs}$ implies that $\frac{d\varrho_{ii}}{dt}=\mathcal{L}_{iiii}$, and since $\varrho_{ii}$  is already at its maximum value, it can only remain constant or decrease, whence $\mathcal{L}_{iiii}\le 0$.
Moreover, we must have that $\mathcal{L}_{jjii}\ge 0$ ($j\neq i$). 
Indeed, if $\varrho_{ii}=1$ and all other components are zero, then we have $\frac{d\varrho_{jj}}{dt}=\mathcal{L}_{jjii}$ ($j\neq i$), and since $\varrho_{jj}=0$ is already at
its minimum value, it can only remain constant or
increase, which proves that $\mathcal{L}_{jjii}\ge 0$.
Both constraints $\mathcal{L}_{iiii}\le 0$, and $\mathcal{L}_{jjii}\ge 0$ ($j\neq i$) can be verified for our evolution equation.

\section{Quantum stochastic thermodynamics}

We now consider 
the previously obtained evolution equation where we add the effect of external forces in order to be able to represent 
work done on the system. Specifically, we introduce the force $\vec{f}_k=(f_{xk},f_{yk},f_{zk})$, applied to the $k$-th particle.
The effect of these forces is described by the potential energy $-\sum_{k=1}^N\vec{f}_k\cdot \vec{r}_k$, and therefore we can write 
\begin{eqnarray}\fl
\nonumber
    \frac{d\varrho}{dt}&=&\frac{1}{i\hbar}\left[\mathcal{H}_0,\varrho\right]-\frac{1}{i\hbar}\sum_{k=1}^{N}\sum_{s=x,y,z}f_{sk}\left[r_{sk},\varrho\right]-\frac{k_B T \beta}{\hbar^2}\sum_{k=1}^{N}m_k\left(\sum_{s=x,y,z}\left[r_{sk},\left[r_{sk},\varrho\right]\right]\right)\\
    \label{maindiffforces}\fl
    &&+\frac{\beta}{2i\hbar}\sum_{k=1}^{N}\left(\sum_{s=x,y,z}\left[r_{sk},\Theta_{sk}\varrho+\varrho\Theta_{sk}\right]\right).
\end{eqnarray}
In studying quantum thermodynamics we have simplified the notation and 
introduced $\varrho=\mathbb{E}\left\lbrace{\rho}\right\rbrace$.
The applied forces should be considered as perturbations to the system. This means that when $\vec{f}_k=0\,,\forall k=1,...,N,$ we have that $ \lim_{t\to\infty}\varrho=\varrho_{eq}=\frac{1}{Z_{qu}}e^{-\frac{\mathcal{H}_0}{k_BT}}$.
Based on this evolution equation for the density matrix, we 
can obtain expressions of the first and second laws of thermodynamics.

\subsection{First law of thermodynamics}

To develop the first law of the thermodynamics, we introduce the internal energy $\mathcal{E}$ of the system, defined as the average value of the Hamiltonian operator, i.e. $\mathcal{E}=\mbox{Tr}(\mathcal{H}_0\varrho)$. 
The time variation of this internal energy can be developed as follows
\begin{eqnarray}\fl
\frac{d\mathcal{E}}{dt}&=&\frac{d\mbox{Tr}(\mathcal{H}_0\varrho)}{dt}=\mbox{Tr}\left(\mathcal{H}_0\frac{d\varrho}{dt}\right).
\end{eqnarray}
We need to substitute here the
terms coming from Eq. (\ref{maindiffforces}). 
The classical Liouvillian term leads to
\begin{equation}\fl
\mbox{Tr}\left(\mathcal{H}_0\left[\mathcal{H}_0,\varrho\right]\right)=\mbox{Tr}\left(\mathcal{H}_0^2\varrho-\mathcal{H}_0\varrho\mathcal{H}_0\right)=0,
\end{equation}
because of the cyclic property of the trace.
Concerning the effect of the forces, we need to develop $\mbox{Tr}\left(\mathcal{H}_0\left[r_{sk},\varrho\right]\right)=\mbox{Tr}\left(\mathcal{H}_0r_{sk}\varrho-\mathcal{H}_0\varrho r_{sk}\right)=\mbox{Tr}\left(\left[\mathcal{H}_0,r_{sk}\right]\varrho\right)$, where we used again the cyclic property of the trace. We can now use the previously obtained relation $[r_{sk},\mathcal{H}_0]=\frac{i\hbar}{m_k}p_{sk}$, and we get \begin{equation}\fl
\mbox{Tr}\left(\mathcal{H}_0\left[r_{sk},\varrho\right]\right)=-\frac{i\hbar}{m_k}\mbox{Tr}\left(p_{sk}\varrho\right)=-i\hbar\mathbb{E}\left\lbrace{v_{sk}}\right\rbrace,
\end{equation}
where $v_{sk}=p_{sk}/m_k$ is the particle velocity component ($s=x,y,z$, $k=1,...,N$). 
An arbitrary noise term delivers the contribution 
\begin{eqnarray}\fl
\nonumber
\mbox{Tr}\left(\mathcal{H}_0\left[r_{sk},\left[r_{sk},\varrho\right]\right]\right)&=&\mbox{Tr}\left(\left[\mathcal{H}_0,r_{sk}\right]\left[r_{sk},\varrho\right]\right)=-\frac{i\hbar}{m_k}\mbox{Tr}\left(p_{sk}\left[r_{sk},\varrho\right]\right)\\
&=&-\frac{i\hbar}{m_k}\mbox{Tr}\left(\left[p_{sk},r_{sk}\right]\varrho\right)=-\frac{\hbar^2}{m_k}.
\end{eqnarray}
Moreover, the friction contribution corresponds to the term 
\begin{eqnarray}\fl
\nonumber
\mbox{Tr}\left(\mathcal{H}_0\left[r_{sk},\Theta_{sk}\varrho+\varrho\Theta_{sk}\right]\right)&=&\mbox{Tr}\left(\left[\mathcal{H}_0,r_{sk}\right]\left(\Theta_{sk}\varrho+\varrho\Theta_{sk}\right)\right)=-\frac{i\hbar}{m_k}\mbox{Tr}\left(p_{sk}\left(\Theta_{sk}\varrho+\varrho\Theta_{sk}\right)\right)\\
&=&-\frac{i\hbar}{m_k}\mathbb{E}\left\lbrace{p_{sk}\Theta_{sk}+\Theta_{sk}p_{sk}}\right\rbrace.
\end{eqnarray}
Summing up all contributions, we obtain the following result
\begin{eqnarray}
\label{first1}
\fl\frac{d\mathcal{E}}{dt}=\sum_{k=1}^{N}\sum_{s=x,y,z}f_{sk}\mathbb{E}\left\lbrace{v_{sk}}\right\rbrace+2\beta\left[\frac{3}{2}N k_B T-\sum_{k=1}^{N}\sum_{s=x,y,z}\frac{1}{2m_k}\mathbb{E}\left\lbrace{\frac{p_{sk}\Theta_{sk}+\Theta_{sk}p_{sk}}{2}}\right\rbrace\right],
\end{eqnarray}
or equivalently
\begin{eqnarray}\fl
\nonumber
\frac{d\mathcal{E}}{dt}&=&\sum_{k=1}^{N}\vec{f}_{k}\cdot\mathbb{E}\left\lbrace{\vec{v}_{k}}\right\rbrace+2\beta\left[\frac{3}{2}N k_B T-\sum_{k=1}^{N}\frac{1}{2m_k}\mathbb{E}\left\lbrace{\frac{\vec{p}_{k}\cdot\vec{\Theta}_{k}+\vec{\Theta}_{k}\cdot\vec{p}_{k}}{2}}\right\rbrace\right]\\
\label{first2}\fl
    &=&\frac{d\mathbb{E}\left\lbrace L\right\rbrace}{dt}+\frac{d\mathbb{E}\left\lbrace Q\right\rbrace}{dt}.
\end{eqnarray}
This represents the first law of the thermodynamics, where we can identify the rate of average work $\frac{d\mathbb{E}\left\lbrace L \right\rbrace}{dt}$ done on the system with the average power $\sum_{k=1}^{N}\vec{f}_{k}\cdot\mathbb{E}\left\lbrace{\vec{v}_{k}}\right\rbrace$ of the external forces. This term is identical to the one obtained in the classical analysis of the problem.

The second term in Eq. (\ref{first2}) represents the  average heat rate $\frac{d\mathbb{E}\left\lbrace Q \right\rbrace}{dt}$ entering the system.
The term $\frac{1}{2m_k}\mathbb{E}\left\lbrace{\frac{\vec{p}_{k}\cdot\vec{\Theta}_{k}+\vec{\Theta}_{k}\cdot\vec{p}_{k}}{2}}\right\rbrace$ represents a kind of modified average kinetic energy of the particle, where we observe a quadratic form (symmetrized) composed of the momentum of the particle and its  friction operator. It is the quantum counterpart of the classical kinetic term $\frac{1}{2m_k}\mathbb{E}\left\lbrace{\vec{p}_{k}\cdot\vec{p}_{k}}\right\rbrace$.
It is important to note that when the modified kinetic energy of the system is {\it smaller} than $\frac{3}{2}N k_B T$ then heat {\it enters} the system, and when the modified kinetic energy is {\it larger} than $\frac{3}{2}N k_B T$ then heat {\it leaves} the system.  This exactly represents the concept of relaxation toward thermodynamic equilibrium.
It is also seen how the friction coefficient $\beta$ governs the rate of convergence to equilibrium that is, the rate at which equipartition is attained.

Our expression of heat flow entering the system allows us to obtain an interesting form of the quantum equipartition theorem.
Indeed, assuming we are in the case with no applied forces, when the thermal equilibrium is reached we have that $\frac{d\mathcal{E}}{dt}=0$, and thus we get the relation
\begin{equation}\fl
    \frac{1}{2m_k}\mathbb{E}\left\lbrace{\frac{\vec{p}_{k}\cdot\vec{\Theta}_{k}+\vec{\Theta}_{k}\cdot\vec{p}_{k}}{2}}\right\rbrace=\frac{3}{2} k_B T,
\end{equation}
where the average value must be determined with the asymptotic canonical distribution
$ \varrho_{eq}=\frac{1}{Z_{qu}}e^{-\frac{\mathcal{H}_0}{k_BT}}$ (and where the subscript $k$ is not summed).
By considering a single component we can write 
\begin{equation}\fl
    \frac{1}{2m_k}\mathbb{E}\left\lbrace{\frac{p_{sk}\Theta_{sk}+\Theta_{sk}p_{sk}}{2}}\right\rbrace=\frac{k_BT}{2},
    \label{equione}
\end{equation}
where $s$ and $k$ are not summed.
The latter relation represents the quantum equipartition theorem written for a single quadratic term of the modified kinetic energy.
This result has now been obtained from the asymptotic behavior of the evolution equation of the density matrix. It can be also proved independently of the evolution equation of the density matrix as follows. 
Assuming that we are at equilibrium, we search for a proof of Eq. (\ref{equione}) when projected onto the energy basis of the system. In this situation, it takes the following explicit form
\begin{equation}\fl
    \frac{1}{2m_kZ_{qu}}\mbox{Tr}\left\lbrace{\frac{p_{sk}\Theta_{sk}+\Theta_{sk}p_{sk}}{2}e^{-\frac{\mathcal{H}_0}{k_BT}}}\right\rbrace=\frac{k_BT}{2},
\end{equation}
or, equivalently,
\begin{equation}\fl
    \frac{1}{2m_k}\sum_r\sum_t p_{sk,rt}p_{sk,tr}\frac{\tanh\left(\frac{E_t-E_r}{2k_B T}\right)}{\frac{E_t-E_r}{2k_B T}}\frac{e^{-\frac{E_i}{k_BT}}}{Z_{qu}}=\frac{k_BT}{2},
    \label{equitwo}
\end{equation}
where we explicitly indicated the sums over $r$ and $t$ (since $s$ and $k$ are not summed). 
Here, as before, we can introduce the quantity $e_n=e^{-\frac{E_n}{k_BT}}>0$, and we can then  write
\begin{equation}\fl
    \frac{1}{2m_k}\sum_r\sum_t\left( p_{sk,rt}p_{sk,tr}\frac{\frac{e_r-e_t}{e_t+e_r}}{\frac{E_t-E_r}{2k_B T}}\right)\frac{e_r}{\sum_k e_k}=\frac{k_BT}{2}.
    \label{equithree}
\end{equation}
The quantity in the round bracket, which we refer to as $c_{rt}$, represents a symmetric matrix, thus satisfying  $c_{rt}=c_{tr}$. Hence, we have that $\sum_r\sum_t c_{rt}e_r=\sum_t\sum_r c_{tr}e_t=\sum_r\sum_t c_{rt}e_t$. 
This allows us to say that $\sum_r\sum_t c_{rt}e_r=\frac{1}{2}(\sum_r\sum_t c_{rt}e_r+\sum_r\sum_t c_{rt}e_t)$, and therefore Eq. (\ref{equithree}) simplifies to 
\begin{equation}\fl
    \frac{k_B T}{2m_k}\sum_r\sum_t\left( \frac{p_{sk,rt}p_{sk,tr}}{E_t-E_r}\right)(e_r-e_t)\frac{1}{\sum_k e_k}=\frac{k_BT}{2}.
    \label{equifour}
\end{equation}
Now, the first quantity in round bracket is a skew-symmetric matrix $d_{rt}=-d_{tr}$, and then we have that $\sum_r\sum_t d_{rt}e_r=\sum_t\sum_r d_{tr}e_t=-\sum_r\sum_t d_{rt}e_t$. So, the expression is further simplified as
\begin{equation}\fl
    \frac{k_B T}{m_k}\sum_r\sum_t\left( \frac{p_{sk,rt}p_{sk,tr}}{E_t-E_r}\right)\frac{e_r}{\sum_k e_k}=\frac{k_BT}{2}.
    \label{equifive}
\end{equation}

To complete the demonstration, we verify that the relationship $\sum_t \frac{p_{sk,rt}p_{sk,tr}}{E_t-E_r}=\frac{m_k}{2}$ is true for any value of $r$.
We previously proved that $r_{sk,rt}=\frac{i\hbar}{m_k}\frac{p_{sk,rt}}{E_t-E_r}$, when we adopt the energy representation. 
The skew-symmetric matrix $d_{rt}=\frac{p_{sk,rt}p_{sk,tr}}{E_t-E_r}$ can be therefore rewritten as $d_{rt}=\frac{m_k}{i\hbar}r_{sk,rt}p_{sk,tr}$.
Since this matrix is skew-symmetric, we also have that $d_{rt}=-\frac{m_k}{i\hbar}r_{sk,tr}p_{sk,rt}$.
Then we can write $d_{rt}=\frac{m_k}{2i\hbar}(r_{sk,rt}p_{sk,tr}-r_{sk,tr}p_{sk,rt})$. Now, from the canonical commutator $[r_{sk},p_{sk}]=i\hbar$, we have that $\sum_t(r_{sk,rt}p_{sk,t\ell}-p_{sk,rt}r_{sk,t\ell})=i\hbar\delta_{r\ell}$, which means $\sum_t(r_{sk,rt}p_{sk,tr}-p_{sk,rt}r_{sk,tr})=i\hbar$, $\forall r$. Hence, we can write $\sum_t d_{rt}=\frac{m_k}{2}$, $\forall r$. We have finally proved that $\sum_t \frac{p_{sk,rt}p_{sk,tr}}{E_t-E_r}=\frac{m_k}{2}$ is true for any value of $r$. Therefore, Eq. (\ref{equifive}) is demonstrated and the quantum equipartition theorem is re-obtained with a calculation developed directly at equilibrium, without using the density matrix evolution equation. Importantly, this direct demonstration proves the equipartition of energy for each individual quadratic term, as shown in Eq. (\ref{equione}).

To summarize, we can say that in quantum mechanics the 
expectation value of each quadratic term of the energy at equilibrium is not equal to $\frac{1}{2}k_B T$, but each modified quadratic term (with the friction operator) takes on exactly the value $\frac{1}{2}k_B T$.

\subsection{Second law of thermodynamics}

In this section we develop a balance equation for the von Neumann entropy of the system defined as follows
\begin{equation}\fl
    \mathcal{S}=-k_B\mbox{Tr}(\varrho\ln\varrho)=-k_B\mathbb{E}(\ln\varrho).
    \label{qentropy}
\end{equation}
Let us first justify mathematically the possibility of calculating the logarithm of the density matrix $\varrho$.
Since $\varrho$ is Hermitian and positive definite, we have the spectral decomposition $\varrho=P^{*T}DP$, where $D$ is diagonal with positive elements (the eigenvalues $p_i$ of $\varrho$), and $P$ is unitary ($P^{-1}=P^{*T}$).
We define a diagonal matrix $L$, whose elements are given by the quantities $\ln p_i$. It means that $D=e^L$. 
We have therefore  that $\varrho=P^{*T}DP=P^{*T}e^LP=e^{P^{*T}LP}$, and we can define the logarithm of the density matrix as $\ln\varrho=P^{*T}LP$.
This procedure makes it possible to determine in principle the logarithm of the density matrix at each instant of time. 

Consequently, we can now calculate the time derivative of entropy
\begin{equation}\fl
    \frac{d\mathcal{S}}{dt}=-k_B\mbox{Tr}\left(\frac{d\varrho}{dt}\ln\varrho+\varrho\frac{d}{dt}\ln\varrho\right).
    \label{derentro}
\end{equation}
We observe
that $\frac{d}{dt}\ln\varrho=\varrho^{-1}\frac{d\varrho}{dt}=\frac{d\varrho}{dt}\varrho^{-1}$ only if $\varrho$ and $\frac{d\varrho}{dt}$ commute. 
This is not true in general for the density matrix, so we must proceed differently. 
There are general formulas for calculating the derivative of the logarithm of a matrix, 
however we prefer to use the following direct technique.
Since $\varrho=P^{*T}DP$, and $\ln\varrho=P^{*T}LP$, we have $\varrho\ln\varrho=P^{*T}DPP^{*T}LP=P^{*T}DLP$.
Therefore, we can write
\begin{equation}\fl
    \mathcal{S}=-k_B\mbox{Tr}(\varrho\ln\varrho)=-k_B\sum_i p_i\ln p_i,
\end{equation}
where $\sum_i p_i=1$ since $\mbox{Tr}(\varrho)=1$.
This expression shows the direct relationship between the von Neumann quantum entropy defined in Eq. (\ref{qentropy}) and the Shannon entropy, introduced and largely used in information theory. 
Moreover, this relation allows us to determine the entropy time derivative by considering the eigenvalues of the density matrix. 
We get
\begin{equation}\fl
    \frac{d\mathcal{S}}{dt}=-k_BT\sum_i\left(\frac{dp_i}{dt}\ln p_i+\frac{dp_i}{dt}\right)=-k_B\sum_i\frac{dp_i}{dt}\ln p_i,
    \label{shannon}
\end{equation}
since $\sum_i\frac{dp_i}{dt}=0$, being $\sum_i p_i=1$. We prove now that this expression exactly coincides with the first term in Eq. (\ref{derentro}), and therefore the second term is zero.
To do this, we develop the following calculation 
\begin{eqnarray}\fl
\nonumber
    -k_B\mbox{Tr}\left(\frac{d\varrho}{dt}\ln\varrho\right)&=&-k_B\mbox{Tr}\left[\left(\dot{P}^{*T}DP+P^{*T}\dot{D}P+P^{*T}D\dot{P}\right)P^{*T}LP\right]\\
    &=&-k_B\mbox{Tr}\left[P\dot{P}^{*T}DL+\dot{P}P^{*T}LD+\dot{D}L\right],
    \label{derentrobis}
\end{eqnarray}
where we used the cyclic property of the trace and the fact that $P^{*T}P=PP^{*T}=I$. Now, $LD=DL$ since $D$ and $L$ are diagonal matrices, and $\frac{d}{dt}(PP^{*T})=P\dot{P}^{*T}+\dot{P}P^{*T}=0$ since $P^{*T}P=PP^{*T}=I$. Hence, we obtain that Eq. (\ref{derentrobis}) is given by $-k_B\sum_i\frac{dp_i}{dt}\ln p_i$.
It means that the first term in Eq. (\ref{derentro}) coincides with Eq. (\ref{shannon}).
To conclude, the time derivative of the quantum entropy can be written as
\begin{equation}\fl
    \frac{d\mathcal{S}}{dt}=-k_B\mbox{Tr}\left(\frac{d\varrho}{dt}\ln\varrho\right)=-k_B\sum_i\frac{dp_i}{dt}\ln p_i.
    \label{entrorate}
\end{equation}
We also have the auxiliary result which states that $\mbox{Tr}\left(\varrho\frac{d}{dt}\ln\varrho\right)=0$. 

We can now develop the entropy balance by considering the density matrix evolution equation given in Eq. (\ref{maindiffforces}).
The Hamiltonian contribution leads to the following term
\begin{eqnarray}\fl
\mbox{Tr}\left(\left[\mathcal{H}_0,\varrho\right]\ln\varrho\right)=\mbox{Tr}\left(\mathcal{H}_0\varrho\ln\varrho-\varrho\mathcal{H}_0\ln\varrho\right)=\mbox{Tr}\left(\mathcal{H}_0\left[\varrho,\ln\varrho\right]\right)=0,
\end{eqnarray}
where we applied the cyclic property of trace, and recalled that $\varrho$ commutes with $\ln\varrho$.
The contribution of external forces is also zero, in fact
\begin{eqnarray}\fl
\mbox{Tr}\left(\left[r_{sk},\varrho\right]\ln\varrho\right)=\mbox{Tr}\left(r_{sk}\varrho\ln\varrho-\varrho r_{sk}\ln\varrho\right)=\mbox{Tr}\left(r_{sk}\left[\varrho,\ln\varrho\right]\right)=0.
\end{eqnarray}
To complete the entropy balance, we rewrite Eq. (\ref{maindiffforces}) in the compact form
\begin{equation}\fl
    \frac{d\varrho}{dt}=\frac{1}{i\hbar}\left[\mathcal{H}_0,\varrho\right]+\mathcal{A}\varrho+\mathcal{R}\varrho=\mathcal{L}\varrho,
    \label{denseq}
\end{equation}
where $\mathcal{A}\varrho$ represents the term with the external force, $\mathcal{R}\varrho$ the friction and noise contributions (the quantum Langevin bath), and $\mathcal{L}\varrho$ the sum of all terms in the evolution equation. 
The symbols  $\mathcal{A}$, $\mathcal{R}$, and $\mathcal{L}$ must be interpreted as superoperators acting on the density matrix.
Since the Hamiltonian and external force contributions are zero in the entropic balance, we can write that
\begin{equation}\fl
    \frac{d\mathcal{S}}{dt}=-k_B\mbox{Tr}\left(\mathcal{R}\varrho\ln\varrho\right).
\end{equation}
Similarly, the average heat rate defined in previous Section can be written as
\begin{equation}\fl
    \frac{d\mathbb{E}\left\lbrace Q\right\rbrace}{dt}=\mbox{Tr}\left(\mathcal{R}\varrho\mathcal{H}_0\right)=k_B T\mbox{Tr}\left(\mathcal{R}\varrho\frac{\mathcal{H}_0}{k_B T}\right).
\end{equation}
Therefore, we can write the second law of the thermodynamics as
\begin{eqnarray}\fl
    \frac{d\mathcal{S}}{dt}&=&\frac{1}{T}\frac{d\mathbb{E}\left\lbrace Q\right\rbrace}{dt}+\frac{d\mathcal{S}_p}{dt},
\end{eqnarray}
where
\begin{equation}\fl
    \frac{d\mathcal{S}_p}{dt}=-k_B\mbox{Tr}\left[\mathcal{R}\varrho\left(\ln \varrho+\frac{\mathcal{H}_0}{k_B T}\right)\right].
    \label{entrprod}
\end{equation}
The term $\frac{1}{T}\frac{d\mathbb{E}\left\lbrace Q\right\rbrace}{dt}$ represents the entropy flow, that is, the amount of entropy entering the system due to heat exchange. 
 In other words, it represents the disorder carried by heat. 
It can be positive or negative depending on whether the heat flow is incoming or outgoing. 
The term $\frac{d\mathcal{S}_p}{dt}$, on the other hand, represents the production of entropy, that is the entropic increase due to the irreversibility of the thermodynamic transformation. 
It must always be positive for consistency with classical thermodynamics. 

The entropy production can be further simplified as follows. Starting with the expression of the asymptotic canonical distribution
$ \varrho_{eq}=\frac{1}{Z_{qu}}e^{-\frac{\mathcal{H}_0}{k_BT}}$, we can calculate its logarithm as
$\ln\varrho_{eq}=\left(\ln\frac{1}{Z_{qu}}\right)I-\frac{\mathcal{H}_0}{k_B T}$.
This expression can be easily proved by taking the exponential $\exp\left[\left(\ln\frac{1}{Z_{qu}}\right)I-\frac{\mathcal{H}_0}{k_B T}\right]$, and by observing that $I$ and $\mathcal{H}_0$ commute. 
Indeed, we find that $\exp\left[\left(\ln\frac{1}{Z_{qu}}\right)I-\frac{\mathcal{H}_0}{k_B T}\right]=\exp\left[\left(\ln\frac{1}{Z_{qu}}\right)I\right]\exp\left( -\frac{\mathcal{H}_0}{k_B T}\right)=\varrho_{eq}$, proving what is required.
Then, in Eq. (\ref{entrprod}) we can substitute $\frac{\mathcal{H}_0}{k_B T}=-\ln\varrho_{eq}+\left(\ln\frac{1}{Z_{qu}}\right)I$. 
It is observed that the term $\left(\ln\frac{1}{Z_{qu}}\right)I$ cannot contribute to the entropy production since $\mbox{Tr}\left[\mathcal{R}\varrho\right]=0$,
due to
trace preservation during time evolution. 
These considerations finally lead to the following form of entropy production
\begin{equation}\fl
    \frac{d\mathcal{S}_p}{dt}=k_B\mbox{Tr}\left[\mathcal{R}\varrho\left(\ln \varrho_{eq}-\ln\varrho\right)\right].
    \label{entroprodfin}
\end{equation}
We consider now the system without externally applied force, which is described by the evolution equation $\frac{d\varrho}{dt}=\frac{1}{i\hbar}\left[\mathcal{H}_0,\varrho\right]+\mathcal{R}\varrho=\mathcal{L}\varrho$. From Eq. (\ref{entroprodfin}), we can also obtain the alternative form
\begin{equation}\fl
    \frac{d\mathcal{S}_p}{dt}=k_B\mbox{Tr}\left[\mathcal{L}\varrho\left(\ln \varrho_{eq}-\ln\varrho\right)\right].
    \label{entroprodnew}
\end{equation}
Indeed, we can replace the relaxation operator $\mathcal{R}$ with $\mathcal{L}$ since we have that $\mbox{Tr}\left(\left[\mathcal{H}_0,\varrho\right]\ln\varrho_{eq}\right)=\mbox{Tr}\left(\left[\mathcal{H}_0,\varrho\right]\ln\varrho\right)=0$.
Importantly, it has been shown that quantum equations for the evolution of the density matrix that are in the form of the classical Fokker-Planck equation involve rates of entropy production which are always non-negative during the relaxation toward equilibrium \cite{oliveira2023,oliveira2024}. 
Thus, we can write 
\begin{equation}\fl
\frac{d\mathcal{S}_p}{dt}\ge 0,
\end{equation}
corresponding to the second law of thermodynamics. It will be also proved numerically with some specific examples in a following Section. 
We remark that the  non-negativity of the quantum entropic production is discussed in several works, with different approaches and methodologies \cite{lebowitz1978,trushechkin2018,trushechkin2019,ruelle2001,pillet2002,kosloff2019}. 

To show a further connection with non-equilibrium thermodynamics, we also introduce the Helmholtz free energy $\mathcal{F}=\mathcal{E}-T\mathcal{S}$, and we study its evolution. We recall the internal energy definition $\mathcal{E}=\mbox{Tr}(\mathcal{H}_0\varrho)$, the entropy definition $\mathcal{S}=-k_B\mbox{Tr}(\varrho\ln\varrho)$, and then we get the expression $\mathcal{F}=\mbox{Tr}(\mathcal{H}_0\varrho)+k_BT\mbox{Tr}(\varrho\ln\varrho)$.
We now remember the previously discussed relation $\mathcal{H}_0=-k_BT\ln\varrho_{eq}+k_BT\left(\ln\frac{1}{Z_{qu}}\right)I$, and we obtain
\begin{equation}\fl
    \mathcal{F}=k_BT\mbox{Tr}(\varrho\ln\varrho-\varrho\ln\varrho_{eq})-k_BT\ln Z_{qu}.
    \label{helm}
\end{equation}
This expression gives the time evolution of the free energy in terms of the density matrix during the relaxation process toward equilibrium. 
It is immediately seen that when equilibrium is reached, the free energy takes on the asymptotic value $\mathcal{F}=-k_BT\ln Z_{qu}$, which corresponds to the well-known expression of equilibrium statistical mechanics.
Furthermore, the first part of Eq. (\ref{helm}), evolving over time, can be identified with the relative entropy of $\varrho$ with respect to $\varrho_{eq}$. We can indeed write
\begin{equation}\fl
    \mathcal{F}=k_BT\mathcal{S}(\varrho\vert\varrho_{eq})-k_BT\ln Z_{qu},
\end{equation}
where we introduced the quantum relative entropy of $\varrho_1$ with respect to $\varrho_2$ as
\begin{equation}\fl
    \mathcal{S}(\varrho_1\vert\varrho_2) \equiv \mbox{Tr}\left[\varrho_1\ln\varrho_1-\varrho_1\ln\varrho_2\right].
\end{equation}
It represents a  
distance measure between the states described by the two density matrices $\varrho_1$ and $\varrho_2$ \cite{vedral2002}.
We remember that  
Klein's inequality affirms that the quantum relative entropy $\mathcal{S}(\varrho_1\vert\varrho_2)$ is non-negative, and it is zero if and only if $\varrho_1=\varrho_2$ \cite{klein1931}.
It means that $\mathcal{S}(\varrho\vert\varrho_{eq})\ge 0$, and we immediately see that free energy always takes values larger than its equilibrium value
\begin{equation}\fl
    \mathcal{F}\ge -k_BT\ln Z_{qu}.
\end{equation}
On the other hand, we can also observe that the free energy always decreases as time increases. Indeed
\begin{equation}\fl
    \frac{d\mathcal{F}}{dt}=k_BT\mbox{Tr}\left(\frac{d\varrho}{dt}\ln\varrho-\frac{d\varrho}{dt}\ln\varrho_{eq}\right),
\end{equation}
where we used the property stating that $\mbox{Tr}\left(\varrho\frac{d}{dt}\ln\varrho\right)=0$, for any density matrix $\varrho$. Since $\frac{d\varrho}{dt}=\mathcal{L}\varrho$, comparing with Eq. (\ref{entroprodnew}), we obtain the important relationship
\begin{equation}\fl
    \frac{d\mathcal{F}}{dt}=-\frac{1}{T}\frac{d\mathcal{S}_p}{dt}. 
\end{equation}
Since $\frac{d\mathcal{S}_p}{dt}\ge 0$, we find that $\frac{d\mathcal{F}}{dt}\le 0$, which means that free energy always decreases toward its value corresponding to thermodynamic equilibrium. 
Indeed, we know from non-equilibrium thermodynamics that a negative value of free energy change is a necessary condition for a process to be spontaneous (irreversibility). 
We finally demonstrated the perfect consistency between our quantum expressions and  
macroscopic non-equilibrium thermodynamics.

\section{The secular approximation}
We develop here an approximation for the dynamic equation of the density matrix valid in the case of energy representation.
First, we observe that under this assumption the evolution equation can be written as
\begin{equation}\fl
    \frac{d\varrho_{nm}}{dt}=\frac{1}{i\hbar}\left(E_n-E_m\right)\varrho_{nm}+\mathcal{R}_{nmij}\varrho_{ij},
\end{equation}
where, for simplicity, we do not consider the effects of external forces (see Eq. (\ref{denseq}), with $\mathcal{A}=0$). The relaxation operator can be obtained as
\begin{equation}\fl
\mathcal{R}_{nmij}=-\delta_{mj}\sum_q\Gamma^+_{nqqi}+\Gamma^+_{jmni}+\Gamma^-_{jmni}-\delta_{ni}\sum_p\Gamma^-_{jppm},
\end{equation}
where
\begin{eqnarray}\fl
\Gamma^-_{abcd}=\frac{k_B T \beta}{\hbar^2}\sum_{k=1}^{N}m_k\sum_{s=x,y,z}r_{sk,ab}r_{sk,cd}\frac{e^{\frac{E_a-E_b}{2k_BT}}}{\cosh\left(\frac{E_a-E_b}{2k_BT}\right)},\\
\fl\Gamma^+_{abcd}=\frac{k_B T \beta}{\hbar^2}\sum_{k=1}^{N}m_k\sum_{s=x,y,z}r_{sk,ab}r_{sk,cd}\frac{e^{\frac{E_d-E_c}{2k_BT}}}{\cosh\left(\frac{E_d-E_c}{2k_BT}\right)}.
\end{eqnarray}
This exact explicit form makes it easy to verify the basic constraints mentioned at the end of Section \ref{dissi} concerning the consistency of time evolution with the properties of the density matrix \cite{happer1972,blum1981,bianco}. 
It also allows us to introduce the secular approximation based on decoupling the elements on the main diagonal, i.e. the populations $\varrho_{nn}$, from the so-called coherences $\varrho_{nm},\, n\neq m$.
In this approximation, it is also assumed that the equation of the generic coherence depends only on the coherence itself, thus resulting in a scalar ordinary differential equation.
For these coherences, we obtain
\begin{equation}\fl
    \frac{d\varrho_{nm}}{dt}=\frac{1}{i\hbar}\left(E_n-E_m\right)\varrho_{nm}-\gamma_{nm}\varrho_{nm}\,\,\, (\mbox{with } n\neq m),
\end{equation}
with 
\begin{equation}\fl
\gamma_{nm}=-\mathcal{R}_{nmnm}=\sum_q\Gamma^+_{nqqn}-\Gamma^+_{mmnn}-\Gamma^-_{mmnn}+\sum_p\Gamma^-_{mppm},
\end{equation}
where $n\neq m$.
These equations represent the phenomenon of decoherence, i.e. $\varrho_{nm}\to 0$ (with $n\neq m$) for $t\to \infty$. The density matrix is in fact diagonal for long times on the energy basis.
For the populations on the main diagonal, we get
\begin{equation}\fl
    \frac{d\varrho_{nn}}{dt}=\sum_{i\neq n}W_{ni}\varrho_{ii}-\left(\sum_{q\neq n}W_{qn}\right)\varrho_{nn},
    \label{pauli}
\end{equation}
where 
\begin{equation}\fl
W_{ni}=\Gamma^+_{inni}+\Gamma^-_{inni}=\frac{k_B T \beta}{\hbar^2}\sum_{k=1}^{N}m_k\sum_{s=x,y,z}\vert r_{sk,ni}\vert^2\frac{e^{\frac{E_i-E_n}{2k_BT}}}{\cosh\left(\frac{E_i-E_n}{2k_BT}\right)},
\label{fermi}
\end{equation}
which are real and positive by construction.
The coefficients
$W_{ni}$ are the probabilities of transition from the state $i$ to the state $n$ per
unit time, in the Markov approximation.
Therefore, Eq. (\ref{pauli}) represents a Pauli master equation, and Eq. (\ref{fermi}) the corresponding Fermi's golden rule \cite{blum1981,fujita1961,alicki1977}.
The detailed balance 
\begin{equation}\fl
    \frac{W_{ni}}{W_{in}}=\frac{e^{-\frac{E_n}{k_BT}}}{e^{-\frac{E_i}{k_BT}}}=\frac{\varrho_{eq,nn}}{\varrho_{eq,ii}}
    \label{detailed}
\end{equation}
 is always verified and ensures the asymptotic convergence to the canonical distribution \cite{alexander2011}.
We try to reconstruct the expressions of heat and entropic production in the secular approximation.
The average heat rate  can be obtained as
\begin{equation}\fl
    \frac{d\mathbb{E}\left\lbrace Q\right\rbrace}{dt}=\mbox{Tr}\left(\mathcal{R}\varrho\mathcal{H}_0\right)=\sum_n(\mathcal{R}\varrho)_{nn}E_n,
\end{equation}
and since $\frac{d\varrho_{nn}}{dt}=(\mathcal{R}\varrho)_{nn}$, from Eq.(\ref{pauli}) we get
\begin{eqnarray}\fl
    \frac{d\mathbb{E}\left\lbrace Q\right\rbrace}{dt}&=&\sum_n\left[\sum_{i\neq n}W_{ni}\varrho_{ii}-\left(\sum_{i\neq n}W_{in}\right)\varrho_{nn}\right]E_n=\sum_n\sum_{i\neq n}\left(W_{ni}\varrho_{ii}E_n-W_{in}\varrho_{nn}E_n\right).
    \label{heatratebis}
\end{eqnarray}
Under the assumption that the density matrix is approximately diagonal with the elements described by the Pauli master equation, the definition of the entropy rate in Eq. (\ref{entrorate}) continues to hold with $p_i=\varrho_{ii}$. Therefore, we can write
\begin{eqnarray}\fl
\nonumber
    \frac{d\mathcal{S}}{dt}&=&-k_B\sum_n\frac{d\varrho_{nn}}{dt}\ln \varrho_{nn}=-k_B\sum_n\left[\sum_{i\neq n}W_{ni}\varrho_{ii}-\left(\sum_{i\neq n}W_{in}\right)\varrho_{nn}\right]\ln \varrho_{nn}\\ \fl
    &=&-k_B\sum_n\sum_{i\neq n}\left(W_{ni}\varrho_{ii}\ln \varrho_{nn}-W_{in}\varrho_{nn}\ln \varrho_{nn}\right).
    \label{entroratebis}
\end{eqnarray}
Summing up, the entropy production rate can be obtained as
\begin{eqnarray}\fl
    \frac{d\mathcal{S}_p}{dt}=\frac{d\mathcal{S}}{dt}-\frac{1}{T}\frac{d\mathbb{E}\left\lbrace Q\right\rbrace}{dt},
\end{eqnarray}
where we can use Eqs. (\ref{heatratebis}) and (\ref{entroratebis}). Now, considering that $\varrho_{eq,nn}=\frac{1}{Z_{qu}}e^{-\frac{E_n}{k_BT}}$, we get $E_n=-k_BT\ln Z_{qu}-k_BT\ln\varrho_{eq,nn}$. 
If we introduce this expression into Eq. (\ref{heatratebis}), we eventually obtain the entropic production in the form
\begin{eqnarray}\fl
    \frac{d\mathcal{S}_p}{dt}=k_B\sum_n\sum_{i\neq n}\left(W_{ni}\varrho_{ii}-W_{in}\varrho_{nn}\right)\ln\frac{\varrho_{eq,nn}}{\varrho_{nn}}.
\end{eqnarray}
This quantity can be rewritten as 
\begin{eqnarray}\fl
    \frac{d\mathcal{S}_p}{dt}&=&\frac{1}{2}k_B\sum_n\sum_{i\neq n}\left(W_{ni}\varrho_{ii}-W_{in}\varrho_{nn}\right)\ln\frac{\varrho_{eq,nn}}{\varrho_{nn}}+\frac{1}{2}k_B\sum_n\sum_{i\neq n}\left(W_{in}\varrho_{nn}-W_{ni}\varrho_{ii}\right)\ln\frac{\varrho_{eq,ii}}{\varrho_{ii}},
\end{eqnarray}
where in the second part we simply swapped the dummy index names.
We can now change the signs to the terms in the second row, getting
\begin{eqnarray}\fl
\nonumber
    \frac{d\mathcal{S}_p}{dt}&=&\frac{1}{2}k_B\sum_n\sum_{i\neq n}\left(W_{ni}\varrho_{ii}-W_{in}\varrho_{nn}\right)\ln\frac{\varrho_{eq,nn}}{\varrho_{nn}}+\frac{1}{2}k_B\sum_n\sum_{i\neq n}\left(W_{ni}\varrho_{ii}-W_{in}\varrho_{nn}\right)\ln\frac{\varrho_{ii}}{\varrho_{eq,ii}}\\ \fl
    &=&\frac{1}{2}k_B\sum_n\sum_{i\neq n}\left(W_{ni}\varrho_{ii}-W_{in}\varrho_{nn}\right)\ln\frac{\varrho_{ii}\varrho_{eq,nn}}{\varrho_{nn}\varrho_{eq,ii}}.
\end{eqnarray}
To conclude, we use the detailed balance stated in Eq.(\ref{detailed}), and we get final form
\begin{eqnarray}\fl
    \frac{d\mathcal{S}_p}{dt}=\frac{1}{2}k_B\sum_n\sum_{i\neq n}\left(W_{ni}\varrho_{ii}-W_{in}\varrho_{nn}\right)\ln\frac{W_{ni}\varrho_{ii}}{W_{in}\varrho_{nn}}.
\end{eqnarray}
Importantly, this is the classical Schnakenberg's form for the entropy production rate associated with a Pauli master equation \cite{sch}. 
We then demonstrated the consistency of the secular approximation with this entropic production expression, widely used in systems described by a master equation \cite{peliti}.
We remark that the Schnakenberg's formula is always nonnegative since the term $W_{ni}\varrho_{ii}-W_{in}\varrho_{nn}$ has the same sign of the term $\ln\frac{W_{ni}\varrho_{ii}}{W_{in}\varrho_{nn}}$.
The entropy production rate remains therefore nonnegative also within the secular approximation.

\section{Applications}

In the following, we describe the implementation and the results obtained for the evolution equation of the density matrix for two specific well studied instances: the harmonic oscillator and the infinite potential well; both cases are of theoretical and applied interest (e.g. for photons, phonons or quantum dots). 
They are described by the following Hamiltonian operator
\begin{equation}\fl
    \mathcal{H}_0=\frac{p^2}{2m} + V(x),
\end{equation}
where $m$ is the mass of the particle and $V(x)$ is the corresponding potential.

\subsection{The harmonic oscillator}

\begin{figure*}[t!]
\centering
\includegraphics[width=17cm]{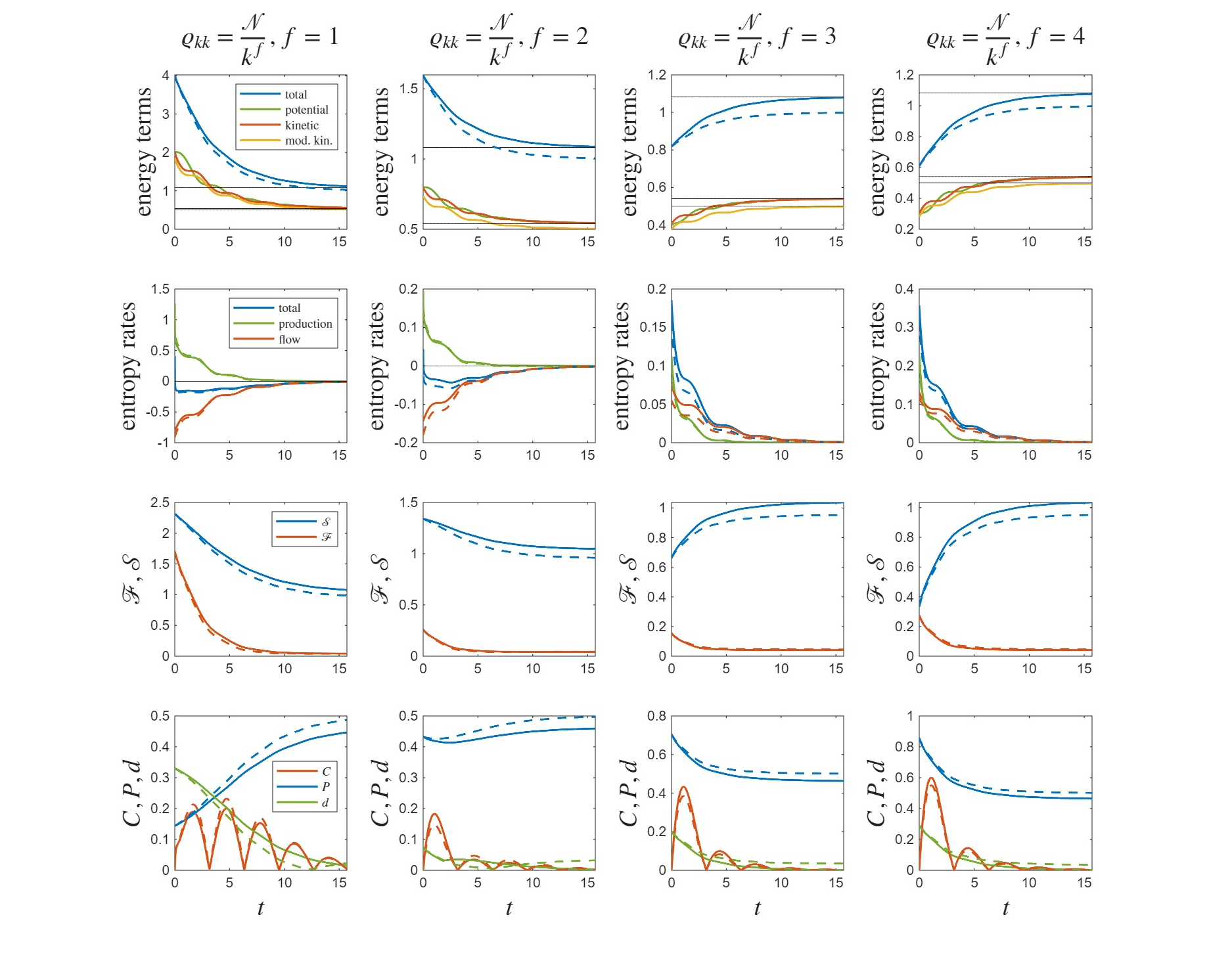}
\caption{\label{oscill}  {Results for the quantum harmonic oscillator in contact with a thermal bath. In each column are represented the plots corresponding to an initial density matrix (16$\times$16) given by $\varrho_{kk}(0)=\mathcal{N}/k^f$, with $f=1,2,3,4$ ($\mathcal{N}$ is a normalizing factor). In the first row, we show the total energy $\mathcal{E}$, the potential energy $\frac{1}{2}m\omega^2\mathbb{E}\left\lbrace x^2\right\rbrace$, the kinetic energy $\frac{1}{2m}\mathbb{E}\left\lbrace p^2\right\rbrace$, and the modified kinetic energy $\frac{1}{2m}\mathbb{E}\left\lbrace{\frac{p\Theta+\Theta p}{2}}\right\rbrace$. We observe that the total energy converges to $\frac{1}{2}\hbar \omega+\hbar \omega/(e^{\frac{\hbar \omega}{k_B T}}-1)$, the potential and kinetic contributions converge to $\frac{1}{4}\hbar \omega+\frac{1}{2}\hbar \omega/(e^{\frac{\hbar \omega}{k_B T}}-1)$, and the modified kinetic energy to $\frac{1}{2}k_B T$, proving the quantum equipartition theorem. In the second row, one can find the behavior of the total entropy rate $\frac{d\mathcal{S}}{dt}$, the entropy flow rate $\frac{1}{T}\frac{d\mathbb{E}\left\lbrace Q\right\rbrace}{dt}$, and the entropy production rate $\frac{d\mathcal{S}_p}{dt}$, which is always positive. In the third row, we plot the entropy $\mathcal{S}$, and the Helmholtz free energy $\mathcal{F}$. We remark that $\mathcal{F}$ is always decreasing and converges to $-k_BT\ln Z_{qu}$.  Finally, in the fourth row, we show the following quantities: the  coherence measure $C=N\Vert\varrho_{wmd}(t)\Vert_2$, where $\varrho_{wmd}$ is $\varrho$ with all diagonal elements turned to zero (without main diagonal), and where $\Vert \cdot\Vert_2$ is the spectral norm; $P$ is the purity measure defined as $\mbox{Tr}(\varrho^2)$ (takes the value 1 if and only if the state is pure); finally, $d$ is the distance between $\varrho$ and $\varrho_{eq}$, defined as $d(t)=\Vert\varrho(t)-\varrho_{eq}\Vert_2$. Dashed lines correspond to the Caldeira-Leggett model. We adopted the parameters $\hbar=1$, $k_B=1$, $T=1$, $\beta=0.3$, $\omega=1$, and $m=1$ in arbitrary units. We calculated 1000 times the exponential of a 256$\times$256 matrix with a time step $\pi/200$.}}
\end{figure*}

In the case of the harmonic oscillator,
we have $V(x) = \frac{1}{2}m\omega^2x^2$ where $\omega=\sqrt{k/m}$ is the classical angular frequency and $k$ is the elastic constant.
We study the relaxation to thermal equilibrium when the harmonic oscillator is embedded in a thermal bath.
We recall that the quantum harmonic oscillator is characterized by the energy levels $\mathcal{H}_0\varphi_{n}=E_n\varphi_{n}$ with 
\begin{equation}\fl
   E_n=\hbar \omega\left(n+\frac{1}{2}\right),\,\,\,\, n\ge 0, 
\end{equation}
and by the eigenfunctions
\begin{equation}\fl
   \varphi_{n}(x)={\frac {1}{\sqrt {2^{n}\,n!}}}\left({\frac {m\omega }{\pi \hbar }}\right)^{1/4}e^{-{\frac {m\omega }{2\hbar }x^{2}}}H_{n}\left({\sqrt {\frac {m\omega }{\hbar }}}x\right), 
\end{equation}
 where $H_n(z)$ are the Hermite polynomials.
We can calculate the matrices associated to the operators $x$ and $p$ as $x_{nm}=\langle \varphi_n(x)\vert x\varphi_m(x) \rangle$ and $p_{nm}=-i\hbar\langle \varphi_n(x)\vert \frac{d}{dx}\varphi_m(x) \rangle$, and we eventually obtain the results
\begin{equation}\fl
 x_{nm}=\sqrt{\frac{\hbar}{2m\omega}}\left(\delta_{n+1,m}\sqrt{n+1}+\delta_{n,m+1}\sqrt{n}\right), 
\end{equation}
and 
\begin{equation}\fl
 p_{nm}=-i\sqrt{\frac{m\omega\hbar}{2}}\left(\delta_{n+1,m}\sqrt{n+1}-\delta_{n,m+1}\sqrt{n}\right), 
\end{equation}
 for $n\ge 0$ and $m\ge 0$.

In the case of the harmonic oscillator, the friction operator defined in Eq. (\ref{frienefind}) is simply obtained as
\begin{eqnarray}\fl
    \Theta_{nm}=p_{nm}\frac{\tanh\left(\frac{\hbar \omega}{2k_B T}\right)}{\frac{\hbar \omega}{2k_B T}},
\end{eqnarray}
and this means that in this particular case $\Theta$ is given by a constant multiplying the operator $p$. 
These premises allow us to write the evolution of the density matrix in the form
\begin{eqnarray}\fl
    \frac{d\varrho}{dt}=\frac{1}{i\hbar}\left[\mathcal{H}_0,\varrho\right]-\frac{k_B T \beta}{\hbar^2}\left[x,\left[x,\varrho\right]\right]+\frac{\beta}{2i\hbar}\frac{\tanh\left(\frac{\hbar \omega}{2k_B T}\right)}{\frac{\hbar \omega}{2k_B T}}\left[x,p\varrho+\varrho p\right].
\end{eqnarray}
We projected that equation onto the energy basis and solved it numerically to observe the thermodynamics of the system 
during its relaxation towards thermal equilibrium.
The results are displayed
in Fig.\ref{oscill}, where we plotted the evolution of main quantities over time for different initial conditions, taking a 16x16 density matrix as a practical example of implementation.
As an initial condition we considered a mixed state composed of the weighted energy eigenfunctions with probability $p_k=\mathcal{N}/k^f$, for chosen values of the parameter $f$, and $\mathcal{N}$ a normalizing factor of the 16$\times$16 density matrix in the energy representation. 
The initial density matrix takes the form $\varrho_{kk}(0)=\mathcal{N}/k^f$, $\varrho_{kh}(0)=0$ if $h\neq k$.
For large values of $f$, the probabilities rapidly become negligible as $k$ increases, and thus only low-energy states are populated; for large values of $f$, the initial energy will be fairly low. 
For small values of $f$, on the other hand, the probabilities decrease more slowly with $k$, and even high-energy states can be appreciably populated. Hence, for small values of $f$, the initial energy can be quite high. 
Recall that the asymptotic mean energy at thermal equilibrium is given by 
$\frac{1}{2}\hbar \omega+\hbar \omega/(e^{\frac{\hbar \omega}{k_B T}}-1)$.
Here we have used the introduced combination of states to obtain two initial conditions in which the initial energy is higher than the thermal equilibrium energy, and two initial conditions in which the initial energy is lower than the thermal equilibrium energy. The first two cases (with $f=1,2$) are shown in the first two columns of Fig.\ref{oscill}, and the other two (with $f=3,4$) in the next two columns. 

In the panels of the first row, we observe the evolution of the energy contributions. 
The total energy decreases towards equilibrium in the first two cases and increases in the following two, consistent with the initial conditions adopted.  
It is well seen that both kinetic and potential contributions converge to the same value corresponding to half of the total equilibrium energy $\frac{1}{2}\hbar \omega+\hbar \omega/(e^{\frac{\hbar \omega}{k_B T}}-1)$. In fact they are both quadratic terms of the Hamiltonian operator.
They do not, however, converge to the classical equipartition value $\frac{k_BT}{2}$, since this principle is no longer strictly verified in the quantum domain. But it can instead be observed that the \textit{modified} kinetic energy $\frac{1}{2m}\mathbb{E}\left\lbrace{\frac{p\Theta+\Theta p}{2}}\right\rbrace$ does indeed converge to the value $\frac{k_BT}{2}$ restoring  the validity of the equipartition theorem, with an extended meaning: the
 difference between $\frac{k_BT}{2}$ and $\frac{1}{2m}\mathbb{E}\left\lbrace{\frac{p\Theta+\Theta p}{2}}\right\rbrace$ during the relaxation describes the heat entering or leaving the system. 

In the first two columns of Fig.\ref{oscill} the total energy decreases because of the outgoing heat and in the other two the energy increases because of the incoming heat. These heat flows generate an entropy flow rate (displacement of disorder) that can be negative or positive.
This can be seen in the panels of the second row of Fig.\ref{oscill}, where the entropy flow rate is negative in the first two cases and positive in the other two. 
The total entropy rate can also have a sign that depends on the initial conditions and the state of progress of the relaxation process. Regardless of the signs of the entropy flow rate and the total entropy rate, it is important to note that the entropy production rate is always positive, in accordance with the second law of thermodynamics. This term, in fact, corresponds to the irreversibility and spontaneity of the process rather than the direction of heat flow.

In the third row of the figure, we depict the evolution of the total entropy $\mathcal{S}$ and the Helmholtz free energy $\mathcal{F}$ as functions of time. While the total entropy may increase or decrease depending on thermal fluxes, the Helmholtz free energy always decreases, eventually reaching its equilibrium value $ -k_BT\ln Z_{qu}=\frac{1}{2}\hbar\omega+k_BT\ln[1-\exp(-\frac{\hbar\omega}{k_BT})]$. 
This decrease characterizes spontaneous processes, consistently with the second law of thermodynamics.

In the last column of Fig. \ref{oscill}, we present several quantities that characterize the evolution of the density matrix. 
Firstly, we consider a measure of coherence, which quantifies the magnitude of the off-diagonal terms in the density matrix. It is defined as $C=N\Vert\varrho_{wmd}(t)\Vert_2$ (with $N$=16 in this example) where $\varrho_{wmd}$ is $\varrho$ with all diagonal elements turned to zero (without main diagonal), and where $\Vert \cdot\Vert_2$ is the spectral norm.
This measure thus describes the decoherence process as the dynamics progresses towards thermodynamic equilibrium.
Note that the system periodically reaches a state without coherence terms (i.e., characterized by a diagonal density matrix), while complete decoherence is achieved only asymptotically due to thermal fluctuations. 
Furthermore, even at points where coherence is absent, the state is never truly pure, as evidenced by the evolution of $P=\mbox{Tr}(\varrho^2)$, which represents the purity of the state ($P=1$ if and only if the state is pure, while $P<1$ for any mixed state).  
It can be observed that the initial state is closer to being pure when $f$ is large (rightmost column), as only a few energy states are occupied. 
Consequently, purity may either increase or decrease during the evolution towards equilibrium, depending on the initial conditions.
Finally, we have represented a quantity that measures the distance $d(t)=\Vert\varrho(t)-\varrho_{eq}\Vert_2$ of the density matrix from its equilibrium value. This quantity is always decreasing, and its decay is comparable to that of coherence.
 {

We introduce a comparison between our model and the Caldeira-Leggett model \cite{caldeira1981,caldeira1983,caldeira1983bis}. Developed in the early 1980s, it was designed to provide a microscopic description of quantum dissipation by coupling a system linearly to a bath of harmonic oscillators that represent its environment. 
This framework enabled the study of quantum Brownian motion. While the model qualitatively captures dissipative dynamics, it presents significant challenges when describing equilibrium statistical properties. 
In particular, the reduced density matrix of the system does not generally yield the canonical Boltzmann distribution expected from statistical mechanics. 
This incompatibility becomes especially pronounced in the strong-coupling or low-temperature regimes, where system-bath correlations are non-negligible. Consequently, the thermodynamic consistency of the model can break down, and naive application of equilibrium concepts may lead to incorrect predictions. 
For the harmonic oscillator, the Caldeira–Leggett model assumes the form
\begin{eqnarray}\fl
    \frac{d\varrho}{dt}=\frac{1}{i\hbar}\left[\mathcal{H}_0,\varrho\right]-\frac{k_B T \beta}{\hbar^2}\left[x,\left[x,\varrho\right]\right]+\frac{\beta}{2i\hbar}\left[x,p\varrho+\varrho p\right].
\end{eqnarray}
Compared with our model, the only difference is that the term $\tanh\left(\frac{\hbar \omega}{2k_B T}\right)/{\frac{\hbar \omega}{2k_B T}}$ does not appear. The results for this model can be observed in Fig. \ref{oscill} and correspond to the dashed lines.
The main point, which highlights the thermodynamic incompatibility, is described by the plots in the first line where we see that the total energy of the system does not asymptotically converge to the expected value, which corresponds to $\frac{1}{2}\hbar \omega+\hbar \omega/(e^{\frac{\hbar \omega}{k_B T}}-1)$.
Although the other curves have a qualitative behavior similar to that of our model, the deviation of the asymptotic regime from the correct one makes the Caldeira Leggett model inconsistent with equilibrium quantum statistical mechanics. This point further justifies the introduction of our approach, based on the canonical quantization of the Fokker-Planck equation, which allows us to correct this drawback and obtain the correct quantum thermodynamics.

It is important to add some comments concerning the numerical solution of the master equations.
If we consider a system with $N$ levels or states, the master equation corresponds to a system of differential equations with $N^2$ unknowns. Since the system is linear, it is described by matrices (or superoperators) having dimensions $N^2\times N^2$.  
The transition from the original master equation to the extended version with $N^2$ unknowns takes place via the introduction of Liouville superoperators, obtained through Kronecker products \cite{campaioli2024}. This is the method adopted in our simulations. Unfortunately, this technique involves extremely large matrices for systems that require a large number of states.
However, for most physical systems the overall Liouvillian superoperator is sparse,
and therefore the number of operations and memory cost associated with its construction and elaboration is reasonably practicable even for certain classes of multi-body systems. 
Of course, there are specific efficient methods for particular systems. 
The covariance matrix approaches are primarily used for Gaussian systems, such as quadratic fermionic/bosonic Hamiltonians, often in the context of Lindblad master equations. For these systems, it is possible to derive equations of motion for the covariance matrix, describing the overall dynamics \cite{cm1,cm2}. However, these techniques can not applicable to strongly interacting, non-Gaussian systems.
Moreover, the Bethe ansatz methods have been developed to solve integrable models in many-body quantum physics (e.g., Heisenberg spin chains), and extended to non-equilibrium master equations, especially in interacting particle systems.
These techniques use the integrability of the system to construct exact eigenstates of the non-Hermitian superoperator governing the master equation \cite{bethe1,bethe2}.
These approaches are only mentioned for completeness but go beyond what is developed in this paper. Moreover, they are often applied to systems where it is not interesting to observe asymptotic behavior for long times, which is instead the main focus of our work.

}
\subsection{The infinite potential well}

\begin{figure*}[t!]
\centering
\includegraphics[width=17cm]{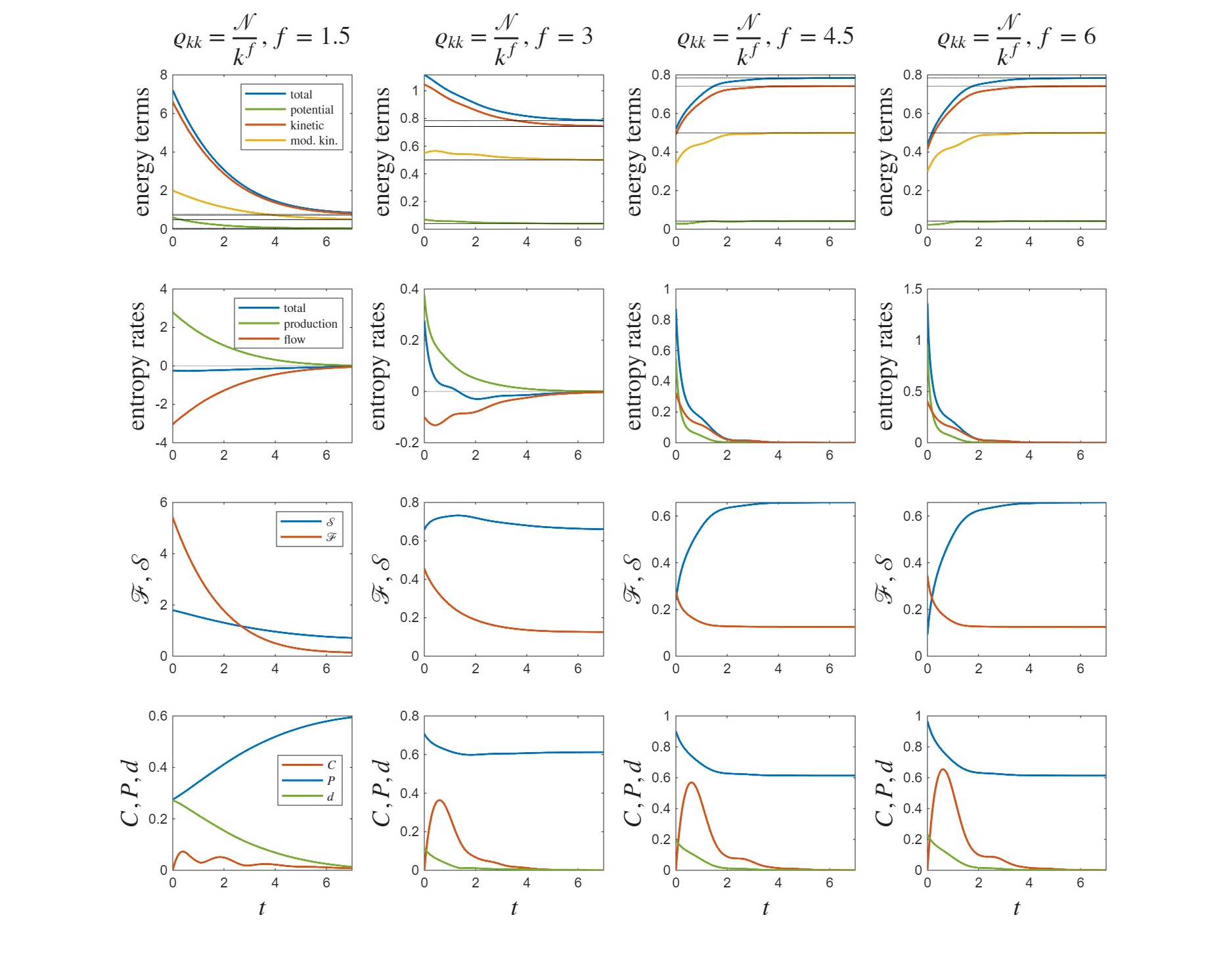}
\caption{\label{inwell} Results for the quantum infinite potential well in contact with a thermal bath. In each column are represented the plots corresponding to an initial density matrix (15$\times$15) given by $\varrho_{kk}(0)=\mathcal{N}/k^f$, with $f=1.5,3,4.5,6$ ($\mathcal{N}$ is a normalizing factor). In the first row, we show the total energy $\mathcal{E}$, the potential energy $\mathcal{E}-\frac{1}{2m}\mathbb{E}\left\lbrace p^2\right\rbrace$, the kinetic energy $\frac{1}{2m}\mathbb{E}\left\lbrace p^2\right\rbrace$, and the modified kinetic energy $\frac{1}{2m}\mathbb{E}\left\lbrace{\frac{p\Theta+\Theta p}{2}}\right\rbrace$. We observe that the modified kinetic energy converges to $\frac{1}{2}k_B T$, proving the quantum equipartition theorem. In the second row, one can find the behavior of the total entropy rate $\frac{d\mathcal{S}}{dt}$, the entropy flow rate $\frac{1}{T}\frac{d\mathbb{E}\left\lbrace Q\right\rbrace}{dt}$, and the entropy production rate $\frac{d\mathcal{S}_p}{dt}$, which is always positive. In the third row, we plot the entropy $\mathcal{S}$, and the Helmholtz free energy $\mathcal{F}$. We remark that $\mathcal{F}$ is always decreasing and converges to $-k_BT\ln Z_{qu}$.  Finally in the fourth row, we show the following quantities: the  coherence measure $C=N\Vert\varrho_{wmd}(t)\Vert_2$, where $\varrho_{wmd}$ is $\varrho$ with all diagonal elements turned to zero (without main diagonal), and where $\Vert \cdot\Vert_2$ is the spectral norm; $P$ is the purity measure defined as $\mbox{Tr}(\varrho^2)$ (takes the value 1 if and only if the state is pure); finally, $d$ is the distance between $\varrho$ and $\varrho_{eq}$, defined as $d(t)=\Vert\varrho(t)-\varrho_{eq}\Vert_2$. We adopted the parameters $\hbar=1$, $k_B=1$, $T=1$, $\beta=1$, $L=2$, and $m=3$ in arbitrary units. We calculated 1000 times the exponential of a 225$\times$225 matrix with a time step $0.007$.}
\end{figure*}

For the infinite potential well the potential energy is
defined as  $V(x)=0$ if $0<x<L$, and $V(x)\to +\infty$ if $x<0$ or $x>L$.
The energy levels for this infinite potential well are given by
\begin{equation}\fl
    E_n=\frac{\pi^2\hbar^2n^2}{2mL^2},\,\,\,\, n\ge 1,
\end{equation}
and they are associated with the eigenfunctions
\begin{equation}\fl
    \varphi_{n}(x)=\sqrt{\frac{2}{L}}\sin\left(\frac{n \pi  x}{L}\right),
\end{equation}
 satisfying the eigenvalue equation  $\mathcal{H}_0\varphi_{n}=E_n\varphi_{n}$. As before, we can evaluate the matrices associated to the operators $x$ and $p$ as $x_{nm}=\langle \varphi_n(x)\vert x\varphi_m(x) \rangle$ and $p_{nm}=-i\hbar\langle \varphi_n(x)\vert \frac{d}{dx}\varphi_m(x) \rangle$. The resulting expressions are 
\begin{equation}\fl
 x_{nm}=\frac{4nmL}{\pi^2}\frac{(-1)^m\cos(n\pi)-1}{(m^2-n^2)^2}(1-\delta_{nm})+\frac{L}{2}\delta_{nm},
\end{equation}
and 
\begin{equation}\fl
 p_{nm}=-i\hbar\frac{2nm}{L}\frac{(-1)^m\cos(n\pi)-1}{m^2-n^2}(1-\delta_{nm}),
\end{equation}
for $n\ge 1$ and $m\ge 1$. The matrix mechanics of the infinite square well is discussed in Refs. \cite{prentis2014,vedral2024}. 

For the infinite potential well the potential energy is
defined as  $V(x)=0$ if $0<x<L$, and $V(x)\to +\infty$ if $x<0$ or $x>L$. The friction operator defined in Eq. (\ref{frienefind}) is  obtained as
\begin{eqnarray}\fl
    \Theta_{nm}=p_{nm}\frac{\tanh\left[\frac{\pi^2\hbar^2}{4mL^2k_B T}(n^2-m^2)\right]}{\frac{\pi^2\hbar^2}{4mL^2k_B T}(n^2-m^2)}.
\end{eqnarray}
These explicit calculations allow us to write the evolution of the density matrix in the form
\begin{equation}\fl
    \frac{d\varrho}{dt}=\frac{1}{i\hbar}\left[\mathcal{H}_0,\varrho\right]-\frac{k_B T \beta}{\hbar^2}\left[x,\left[x,\varrho\right]\right]+\frac{\beta}{2i\hbar}\left[x,\Theta\varrho+\varrho \Theta\right].
\end{equation}
As before, we projected that equation onto the energy basis and solved it numerically to observe the thermodynamics  during the relaxation toward thermal equilibrium.
The results can be found in Fig. \ref{inwell},  where we plotted  the  evolution  of  main  quantities  over  time  for  different initial  conditions. As before, we adopted two initial conditions corresponding to an initial energy larger than the equilibrium energy (first two columns), and two initial conditions with energy smaller than its equilibrium value (following two columns).  
It can be seen from the graphs in the first row that the total energy converges to the thermodynamic value $\mathcal{E}=\frac{1}{Z_{qu}}\sum_{n=1}^{+\infty}E_n\exp(-\frac{E_n}{k_BT})$, where $Z_{qu}=\sum_{n=1}^{+\infty}\exp(-\frac{E_n}{k_BT})$. 
In this case the kinetic and potential contributions do not converge to the same limit because the potential term is not quadratic. 
The former converges toward $\frac{1}{2m}\mathbb{E}\left\lbrace p^2\right\rbrace$ and the latter towards $\mathcal{E}-\frac{1}{2m}\mathbb{E}\left\lbrace p^2\right\rbrace$, where the average value is calculated through the canonical distribution.
In addition, the modified kinetic energy converges towards $\frac{1}{2}k_B T$, numerically proving the new version of the energy equipartition.
In the second row of plots we see that the entropy flow rate is negative or positive depending on whether the heat is outgoing or incoming. 
The total entropy rate can also be positive or negative depending on the direction of heat flow. 
But entropy production is seen to be always positive since it is related to the irreversibility of the relaxation process.
In the third line we show the evolution of entropy and Helmholtz free energy: the total entropy can increase or decrease while the free energy always decreases until it reaches the equilibrium value $ -k_BT\ln Z_{qu}$.
In the fourth and final row of Fig. \ref{inwell}, we show the evolution of the coherence measure, of the purity measure and of the distance from the canonical distribution.
It is observed that while starting from a diagonal density matrix the coherence first increases and then decreases going towards thermodynamic equilibrium. The purity increases or decreases depending on the initial condition, as already observed for the harmonic oscillator. Finally, the distance of the density matrix from the canonical distribution is monotonically decreasing with time. 

\section{Conclusions}

In this work, we have exploited the formal connections between classical stochastic thermodynamics and quantum open systems, with a particular focus on the interplay between thermal noise and dissipation. 
We initially proved that thermal noise in quantum mechanics is not compatible with Schrödinger equation since the state cannot be pure. 
We then introduced the mixed states via the density matrix described by the Liouville-von Neumann equation. 
The quantum effects of thermal noise are proved to be equivalent to those of a multi-dimensional geometric Brownian process. 
This enabled the introduction of appropriate terms describing thermal fluctuations into the master equation and they are closely resembling to those classically obtained. 
By further leveraging analogies with classical equations, we introduced a novel Hermitian dissipation operator, which enables a reasonable quantum description of friction. 
This operator provides a clear physical interpretation of dissipation within quantum systems.

The new friction operator allows for a physically clear definition of heat exchange and therefore leads to the introduction of quantum thermodynamics. 
Indeed, we demonstrated that the proposed framework allows for a well-defined formulation of the first and second laws of thermodynamics in the quantum regime. 
An alternative  \textit{quantum equipartition theorem}, derived from our approach, offers a new perspective on the distribution of energy in open quantum systems, distinguishing it from classical equipartition results.
In classical equipartition, each quadratic energy term corresponds at equilibrium to an energy contribution equal to $\frac{k_BT}{2}$. 
This is not true in quantum mechanics, and so one must change either the definition of kinetic energy, or the equilibrium value $\frac{k_BT}{2}$. 
While various works in the literature have investigated the new value  to substitute in place of $\frac{k_BT}{2}$, here we have modified the definition of kinetic energy so as to respect the classical value. 
This can be done precisely because of the mathematical form of the new friction operator. 
This result allows us to provide a new contribution to the many studies of quantum equipartition found in the recent literature \cite{bialas2018,spiechowicz2018,bialas2019,luczka2020,tong2024}.

Our formalism ensures that energy dissipation and entropy production adhere to fundamental thermodynamic principles, reinforcing the consistency between quantum and classical descriptions of non-equilibrium processes. 
In particular,  the entropy production rate is always positive, confirming the validity of the second law.
The proposed model therefore represents a useful approximation describing the quantum Brownian motion of an arbitrary system. It is particularly useful for modelling mesoscopic systems with an intermediate behaviour between classical and quantum regimes. Additionally, our results highlight the intricate role of quantum fluctuations in shaping the thermodynamic behavior of quantum systems, which has implications for the stability and control of quantum states in practical applications.

As a first, simple validation of our theoretical framework, we applied it to two fundamental quantum systems: the harmonic oscillator and a particle in an infinite potential well. Our numerical analyses confirmed the expected relaxation dynamics and thermodynamic behavior, including the irreversible approach to thermal equilibrium. The results underscore the robustness of our formulation in capturing the essential features of quantum dissipation and stochastic dynamics. Furthermore, the application of our model to these well-defined quantum systems provides a benchmark for testing future developments in quantum thermodynamics, particularly in scenarios where quantum irreversibility and decoherence effects become relevant. We also discussed a comparison with the Caldeira-Leggett model.

Beyond its immediate applications, this work provides a foundation for further research in quantum thermodynamics, stochastic quantum processes, and the development of quantum technologies. 
Future studies could extend this framework to more complex quantum systems, including many-body interactions and strongly correlated environments. Additionally, the incorporation of non-Markovian effects  into our model could yield deeper insights into quantum thermodynamics and its applications in emerging quantum technologies. 
Understanding the interplay between dissipation and decoherence is crucial for advancing quantum computing, quantum sensing, and the design of novel quantum devices that operate far from equilibrium.

Our results also contribute to the ongoing efforts to establish a unified description of non-equilibrium processes in both classical and quantum domains. 
The introduction of a Hermitian dissipation operator opens new possibilities for refining the theoretical underpinnings of quantum thermodynamics, with potential implications for quantum nanotechnologies. 
Moreover, our findings pave the way for experimental verification, as the proposed framework can be tested in controlled quantum systems such as trapped ions, superconducting qubits, and ultracold atomic gases. 
Future experimental studies could provide valuable feedback for refining theoretical models, ultimately bridging the gap between abstract quantum thermodynamics and its real-world implementations.

\appendix

\section{Multi-dimensional geometric Brownian motion}
\label{appa}

We consider a stochastic differential equation 
\begin{equation}\fl
    \frac{d\vec{y}}{dt}=\left(C+\sum_{j=1}^\mathfrak{m}D_jn_j(t)\right)\vec{y},
    \label{app1}
\end{equation}
where $\vec{y}\in\mathbb{C}^\mathfrak{n}$, $C$ and $D_j$ are arbitrary complex matrices $\mathfrak{n}\times\mathfrak{n}$, and the real Gaussian noises $n_j(t)$ ($\forall j = 1, ..., \mathfrak{m}$)  satisfy the properties $\mathbb{E}\{  n_j(t)\} = 0$, and $\mathbb{E}\{ n_i(t_1)  n_j(t_2)\} = 2 \delta_{ij} \delta(t_1-t_2) $. 
We need to determine the average value $\mathbb{E}\left\lbrace\vec{y}\right\rbrace$ of the vector $\vec{y}\in\mathbb{C}^\mathfrak{n}$. 
In order to use the Fokker-Planck equation stated in Eq. (\ref{fokpla}), we need to move from complex to real variables. 
To begin, we consider a simple map $\vec{z}=A\vec{w}:\mathbb{C}^\mathfrak{n}\to\mathbb{C}^\mathfrak{n}$, where $A$ is an arbitrary complex matrix $\mathfrak{n}\times\mathfrak{n}$.
If $A=R+iM$, with $R$ and $M$ real matrices $\mathfrak{n}\times\mathfrak{n}$, the decomplexification procedure can be written as
\begin{equation}\fl
    \left(\begin{array}{l}
        {\mathbb{R}\mbox{e}\,}\vec{z}\\ {\mathbb{I}\mbox{m}\,}\vec{z}
    \end{array}\right)=
    \left(\begin{array}{cc}
        R&-M\\ M&R
    \end{array}\right)\left(\begin{array}{l}
        {\mathbb{R}\mbox{e}\,}\vec{w}\\ {\mathbb{I}\mbox{m}\,}\vec{w}
    \end{array}\right)=\tilde A\left(\begin{array}{l}
        {\mathbb{R}\mbox{e}\,}\vec{w}\\ {\mathbb{I}\mbox{m}\,}\vec{w}
    \end{array}\right),
    \label{tildef}
\end{equation}
where $\tilde A$ is a real matrix $2\mathfrak{n}\times 2\mathfrak{n}$.
We verify that this operation does not substantially change the spectrum of the matrix. We introduce the matrix $U$ and its complex conjugate as
\begin{equation}\fl
U=\frac{1}{\sqrt{2}}    \left(\begin{array}{cc}
        I_\mathfrak{n}&iI_\mathfrak{n}\\ iI_\mathfrak{n}&I_\mathfrak{n}
    \end{array}\right)\Rightarrow U^*=\frac{1}{\sqrt{2}}    \left(\begin{array}{cc}
        I_\mathfrak{n}&-iI_\mathfrak{n}\\ -iI_\mathfrak{n}&I_\mathfrak{n}
    \end{array}\right),
\end{equation}
where $I_\mathfrak{n}$ is the identity matrix of order $\mathfrak{n}$. It is possible to verify that $UU^*=I_{2\mathfrak{n}}$, or equivalently $U^{-1}=U^*$. Now we can also verify that
\begin{equation}\fl
    U\tilde A U^*=U    \left(\begin{array}{cc}
        R&-M\\ M&R
    \end{array}\right)U^*=\left(\begin{array}{cc}
        R+iM&0\\ 0&R-iM
    \end{array}\right),
\end{equation}
and similarly
\begin{equation}\fl
    U
    \left(\begin{array}{cc}
        R-\lambda I_\mathfrak{n}&-M\\ M&R-\lambda I_\mathfrak{n}
    \end{array}\right)U^*=\left(\begin{array}{cc}
        R+iM-\lambda I_\mathfrak{n}&0\\ 0&R-iM-\lambda I_\mathfrak{n}
    \end{array}\right).
\end{equation}
By taking the determinant of both sides, we get
\begin{equation}\fl
    \det    \left(\tilde A-\lambda I_{2\mathfrak{n}}\right)=\det(
        R+iM-\lambda I_\mathfrak{n})\det(R-iM-\lambda I_\mathfrak{n}).
\end{equation}
It means that the eigenvalues of $\tilde A$ are the eigenvalues of $A=R+iM$ combined with those of $A^*=R-iM$. In other words, if $\lambda$ is an eigenvalue of $A=R+iM$, then $\lambda$ and $\lambda^*$ are eigenvalues of $\tilde A$.
That said, we can decomplexify Eq. (\ref{app1}) by introducing $\vec{x}=\left(\begin{array}{l}
        {\mathbb{R}\mbox{e}\,}\vec{y}\\ {\mathbb{I}\mbox{m}\,}\vec{y}
    \end{array}\right)\in\mathbb{R}^{2\mathfrak{n}}$, and we get
\begin{equation}\fl
    \frac{d\vec{x}}{dt}=\left(\tilde C+\sum_{j=1}^\mathfrak{m}\tilde D_j n_j(t)\right)\vec{x},
    \label{app2}
\end{equation}
where the  operator $\sim$ is defined in Eq. (\ref{tildef}).
We can now use the Fokker-Planck equation stated in Eq. (\ref{fokpla}), by first defining $h_i=\tilde C_{ik}x_k$, and $g_{ij}=\tilde D_{j,ik}x_k$ (with sums over $k$).
This Fokker-Planck equation assumes the form
\begin{eqnarray}\fl
 \frac{\partial W}{\partial t} = -  \frac{\partial}{\partial x_i} \left[\left(\tilde C_{ik}x_k+2\alpha \tilde D_{j,ks}\tilde D_{j,ik}  x_s \right)W\right] 
 +   \frac{\partial^2 }{\partial x_i \partial x_j}  \left[  \left(\tilde D_{k,it}x_t\tilde D_{k,jp}x_p \right)W \right]  ,
\end{eqnarray}
and allows us to determine the evolution of $\mathbb{E}\left\lbrace x_q\right\rbrace=\int_{\mathbb{R}^{2\mathfrak{n}}} x_q W d\vec{x}$.
We find the time derivative as
\begin{eqnarray}\fl
\nonumber
 \frac{d\mathbb{E}\left\lbrace x_q\right\rbrace}{dt}&=&\int_{\mathbb{R}^{2\mathfrak{n}}}x_q\frac{\partial W}{\partial t}  d\vec{x}=\int_{\mathbb{R}^{2\mathfrak{n}}}-x_q  \frac{\partial}{\partial x_i} \left[\left(\tilde C_{ik}x_k+2\alpha \tilde D_{j,ks}\tilde D_{j,ik}  x_s \right)W\right] d\vec{x}\\
 \nonumber
 \fl&&+\int_{\mathbb{R}^{2\mathfrak{n}}}x_q   \frac{\partial^2 }{\partial x_i \partial x_j}  \left[  \left(\tilde D_{k,it}x_t\tilde D_{k,jp}x_p \right)W \right]  d\vec{x}\\
 \fl&=&\int_{\mathbb{R}^{2\mathfrak{n}}}\delta_{qi}   \left[\left(\tilde C_{ik}x_k+2\alpha \tilde D_{j,ks}\tilde D_{j,ik}  x_s \right)W\right] d\vec{x}=\tilde C_{qk}\mathbb{E}\left\lbrace x_k\right\rbrace+2\alpha \tilde D_{j,ks}\tilde D_{j,qk}  \mathbb{E}\left\lbrace x_s\right\rbrace,
\end{eqnarray}
where we used the property $\int_{\mathbb{R}^{2\mathfrak{n}}}\phi\frac{\partial \psi}{\partial x_i}d\vec{x}=-\int_{\mathbb{R}^{2\mathfrak{n}}}\psi\frac{\partial \phi}{\partial x_i}d\vec{x}$, valid when the functions $\phi$ and $\psi$ are sufficiently regular at infinity.
It means that
\begin{equation}\fl
    \frac{d\mathbb{E}\left\lbrace\vec{x}\right\rbrace}{dt}=\left(\tilde C+2\alpha\sum_{j=1}^\mathfrak{m}\tilde D_j^2\right)\mathbb{E}\left\lbrace\vec{x}\right\rbrace.
\end{equation}
Coming back to the complex notation we finally obtain the ordinary differential equation
\begin{equation}\fl
    \frac{d\mathbb{E}\left\lbrace\vec{y}\right\rbrace}{dt}=\left(C+2\alpha\sum_{j=1}^\mathfrak{m}D_j^2\right)\mathbb{E}\left\lbrace\vec{y}\right\rbrace,
\end{equation}
which is the result used in the main text.
A similar approach can be found in Ref.\cite{palla2020}.

\section{Solution of the Lyapunov equation}
\label{appb}
We consider a matrix equation of the form
\begin{equation}\fl
    AX+XB=C,
    \label{lyap}
\end{equation}
where $A$, $B$, $C$, and $X$ are complex matrices $\mathfrak{n}\times\mathfrak{n}$. 
We can use the property given in Eq. (\ref{kron1}), and we get
\begin{equation}\fl
    \left(A \otimes I-I\otimes B^T\right)\hat{X}=\hat{C}\in\mathbb{C}^{\mathfrak{n}^2}.
    \label{lyapvec}
\end{equation}
This equation can be used for numerical applications when the matrix $A \otimes I-I\otimes B^T$ ($\mathfrak{n}^2\times\mathfrak{n}^2$) is nonsingular. 
We search therefore  for the  $\mathfrak{n}^2$ eigenvalues of the matrix $A \otimes I-I\otimes B^T$ to detect when Eq. (\ref{lyap}) or Eq. (\ref{lyapvec}) has only one solution.
We suppose that $\hat{U}\in\mathbb{C}^{\mathfrak{n}^2}$ is an eigenvector of $A \otimes I-I\otimes B^T$ with eigenvalue $\gamma$. 
We then have
\begin{equation}\fl
    \left(A \otimes I-I\otimes B^T\right)\hat{U}=\gamma\hat{U},
\end{equation}
which is equivalent to 
\begin{equation}\fl
    AU+UB=\gamma U.
\end{equation}
Let $v\in\mathbb{C}^\mathfrak{n}$ now be an eigenvector of $B$ with eigenvalue $\mu$, i.e. $Bv=\mu v$.
We can write
\begin{equation}\fl
    AUv+UBv=\gamma Uv\Rightarrow AUv+\mu Uv=\gamma Uv,
\end{equation}
which corresponds to
\begin{equation}\fl
     A(Uv)=(\gamma-\mu) Uv.
\end{equation}
This equation states that $\gamma-\mu$ is an eigenvalue, say $\lambda$, of $A$.
So, we obtain that $\lambda=\gamma-\mu$, or $\gamma=\lambda+\mu$. Finally, we can state that the eigenvalues $\gamma$ of $A \otimes I-I\otimes B^T$ are given by all the possible sums $\lambda+\mu$ of the eigenvalues of $A$ and $B$.
We finally proved that the matrix problem in Eq. (\ref{lyap}), or Eq. (\ref{lyapvec}), has a single solution if and only if all possible sums of the eigenvalues of matrices $A$ and $B$ are different from zero. 
We consider now the differential problem 
\begin{equation}\fl
    \dot{X}=AX+XB,
    \label{lyapdiff}
\end{equation}
where $A$, $B$, $C$, and $X$ are complex matrices $\mathfrak{n}\times\mathfrak{n}$.
It can be rewritten in vectorized form as
\begin{equation}\fl
    \dot{\hat{X}}=\left(A \otimes I-I\otimes B^T\right)\hat{X}\in\mathbb{C}^{\mathfrak{n}^2},
\end{equation}
with solution
\begin{equation}\fl
    {\hat{X}}(t)=\exp\left[\left(A \otimes I-I\otimes B^T\right)(t-t_0)\right]\hat{X}(t_0).
\end{equation}
For the the property proved above, $\lim_{t\to\infty}{\hat{X}}(t)=0$ if and only if ${\mathbb{R}e}(\lambda_i+\mu_j)<0$ for all $i$ and $j$, where $\lambda_i$ and $\mu_j$ are the eigenvalues of $A$ and $B$, respectively.
Coming back to the matrix notation, we obtain the solution in the form
\begin{equation}\fl
    X(t)=e^{A(t-t_0)}X(t_0)e^{B(t-t_0)},
    \label{lyapdiffsol}
\end{equation}
which can be proved observing that
\begin{eqnarray}\fl
    \dot{X}(t)&=&Ae^{A(t-t_0)}X(t_0)e^{B(t-t_0)}+e^{A(t-t_0)}X(t_0)e^{B(t-t_0)}B=AX+XB.
\end{eqnarray}
From previous results, we deduce that $X(t)=e^{A(t-t_0)}X(t_0)e^{B(t-t_0)}\to 0$, when $t\to\infty$, if and only if ${\mathbb{R}e}(\lambda_i+\mu_j)<0$ for all $i$ and $j$.
If we are in the condition in which
$X(t)\to 0$ when $t\to\infty$, we can write
\begin{equation}\fl
    \int_{t_0}^{+\infty}\dot{X}(t)dt=A\int_{t_0}^{+\infty}X(t)dt+\int_{t_0}^{+\infty}X(t)dtB,
\end{equation}
or
\begin{equation}\fl
    -X(t_0)=A\int_{t_0}^{+\infty}X(t)dt+\int_{t_0}^{+\infty}X(t)dtB.
\end{equation}
If $X(t_0)=C$, and $t_0=0$, we get
\begin{equation}\fl
    -C=A\int_{_0}^{+\infty}e^{At}Ce^{Bt}dt+\int_{_0}^{+\infty}e^{At}Ce^{Bt}dtB.
\end{equation}
We have therefore proved that the equation $AX+XB=C$ has only one solution given by
\begin{equation}\fl
    X=-\int_{_0}^{+\infty}e^{At}Ce^{Bt}dt,
\end{equation}
when ${\mathbb{R}\mbox{e}}(\lambda_i+\mu_j)<0$ for all $i$ and $j$. In particular, the equation  $AX+XA=C$ has only one solution given by
\begin{equation}\fl
    X=-\int_{_0}^{+\infty}e^{At}Ce^{At}dt,
\end{equation}
when ${\mathbb{R}e}(\lambda_i)<0$ for all $i$.

\section{Alternative integral form of the friction operator}
\label{appc}
We start by Eq. (\ref{mainequationb}) combined with Eq. (\ref{quadiff}), resulting in
\begin{eqnarray}\fl
    i\frac{2 m_k k_B T }{\hbar}\left[r_{sk},e^{-\frac{\mathcal{H}_0}{k_BT}}\right]=\Theta_{sk}e^{-\frac{\mathcal{H}_0}{k_BT}}+e^{-\frac{\mathcal{H}_0}{k_BT}}\Theta_{sk},
    \label{mainequationbis}
\end{eqnarray}
and we define the following Heisenberg operators
\begin{eqnarray}\fl
    \tilde{r}_{sk}(t)=e^{+\frac{i\mathcal{H}_0t}{\hbar}}{r}_{sk}e^{-\frac{i\mathcal{H}_0t}{\hbar}},\,\,\,\tilde{\Theta}_{sk}(t)=e^{+\frac{i\mathcal{H}_0t}{\hbar}}{\Theta}_{sk}e^{-\frac{i\mathcal{H}_0t}{\hbar}},\,\,\,\tilde{p}_{sk}(t)=e^{+\frac{i\mathcal{H}_0t}{\hbar}}{p}_{sk}e^{-\frac{i\mathcal{H}_0t}{\hbar}}.
\end{eqnarray}
Therefore, Eq. (\ref{mainequationbis}) can be rewritten as
\begin{eqnarray}\fl
    \gamma_k\left(\tilde{r}_{sk}-e^{-\frac{\mathcal{H}_0}{k_BT}}\tilde{r}_{sk}e^{+\frac{\mathcal{H}_0}{k_BT}}\right)=\tilde{\Theta}_{sk}+e^{-\frac{\mathcal{H}_0}{k_BT}}\tilde{\Theta}_{sk}e^{+\frac{\mathcal{H}_0}{k_BT}},
    \label{mainequationbish}
\end{eqnarray}
where $\gamma_k=i\frac{2 m_k k_B T }{\hbar}$.
We consider now the matrix expression ${f}(\lambda)=e^{\lambda {A}}{B}e^{-\lambda {A}}$, where ${A}$ and ${B}$ are constant matrices and $\lambda$ is a scalar parameter. 
We note that this quantity is the solution of the differential problem $\frac{d{f}(\lambda)}{d\lambda}=[{A},{f}(\lambda)]$, with the initial condition ${f}(0)={B}$, see Eq. (\ref{lyapdiffsol}). We can prove the following Baker-Hausdorff formula
 \begin{equation}\fl
     e^{ {A}}{B}e^{-{A}}={B}+[{A},{B}]+\frac{1}{2!}[{A},[{A},{B}]]+\frac{1}{3!}[{A},[{A},[{A},{B}]]]+...
     \label{baker}
 \end{equation}
Indeed, we can develop ${f}(\lambda)=e^{\lambda {A}}{B}e^{-\lambda {A}}$ in power series as ${f}(\lambda)=\sum_{k=0}^{+\infty}\frac{{f}^{(k)}(0)}{k!}\lambda^k$, and we can obtain the derivatives by starting from ${f}(0)={B}$, and $\frac{d{f}(\lambda)}{d\lambda}=[{A},{f}(\lambda)]$. Iteratively, we obtain that $\frac{d^2{f}(\lambda)}{d\lambda^2}=[{A},[{A},{f}(\lambda)]]$, and similar expressions for higher orders.
 Then, we have ${f}(0)={B}$, $\frac{d{f}(0)}{d\lambda}=[{A},{B}]$, $\frac{d^2{f}(0)}{d\lambda^2}=[{A},[{A},{B}]]$, and so on. When we substitute $\lambda=1$ in the power series, we obtain Eq. (\ref{baker}). This result can be used in both left and right hand sides of Eq. (\ref{mainequationbish}), where we can also introduce the relations $\frac{d}{dt}\tilde{r}_{sk}=\frac{1}{i\hbar}[\tilde{r}_{sk},\mathcal{H}_0]$, and $\frac{d}{dt}\tilde{\Theta}_{sk}=\frac{1}{i\hbar}[\tilde{\Theta}_{sk},\mathcal{H}_0]$, typical of the Heisenberg picture. So doing, the Baker-Hausdorff series become Taylor expansions, and we get
\begin{equation}\fl
    \tilde{\Theta}_{sk}(t)+\tilde{\Theta}_{sk}\left(t+i\frac{\hbar}{k_BT}\right)=\gamma_k\left[\tilde{r}_{sk}(t)-\tilde{r}_{sk}\left(t+i\frac{\hbar}{k_BT}\right)\right],
    \label{comptime}
\end{equation}
where we have implied an analytic continuation from real to complex times. We can now introduce the Fourier transform of a function $g(t)$ as $\mathcal{F}\left\lbrace g\right\rbrace(\omega)=\int_{-\infty}^{+\infty}g(t)e^{-i\omega t}dt$. By Fourier transforming Eq. (\ref{comptime}), we obtain
\begin{equation}\fl
    \mathcal{F}\left\lbrace \tilde{\Theta}_{sk}\right\rbrace(\omega)\left(1+e^{-\frac{\omega\hbar}{k_BT}}\right)=\gamma_k\mathcal{F}\left\lbrace \tilde{r}_{sk}\right\rbrace(\omega)\left(1-e^{-\frac{\omega\hbar}{k_BT}}\right),
    \label{comptimef}
\end{equation}
from which we get
\begin{eqnarray}\fl
    \mathcal{F}\left\lbrace \tilde{\Theta}_{sk}\right\rbrace(\omega)&=&\gamma_k\mathcal{F}\left\lbrace \tilde{r}_{sk}\right\rbrace(\omega)\frac{1-e^{-\frac{\omega\hbar}{k_BT}}}{1+e^{-\frac{\omega\hbar}{k_BT}}}=\gamma_k\mathcal{F}\left\lbrace \tilde{r}_{sk}\right\rbrace(\omega)\tanh\left(\frac{\omega\hbar}{2k_BT}\right).
    \label{comptimefsol}
\end{eqnarray}
By using the definition of $\gamma_k$, this equation can be rewritten as
\begin{eqnarray}\fl
    \mathcal{F}\left\lbrace \tilde{\Theta}_{sk}\right\rbrace(\omega)&=&i\omega m \mathcal{F}\left\lbrace \tilde{r}_{sk}\right\rbrace(\omega)\frac{\tanh\left(\frac{\omega\hbar}{2k_BT}\right)}{\frac{\omega\hbar}{2k_BT}}=\mathcal{F}\left\lbrace \tilde{p}_{sk}\right\rbrace(\omega)\frac{\tanh\left(\frac{\omega\hbar}{2k_BT}\right)}{\frac{\omega\hbar}{2k_BT}},
    \label{comptimefsolbis}
\end{eqnarray}
where we used the relation $\frac{d}{dt}\tilde{r}_{sk}=\frac{1}{i\hbar}[\tilde{r}_{sk},\mathcal{H}_0]=\frac{1}{m}\tilde{p}_{sk}$, corresponding to $i\omega m \mathcal{F}\left\lbrace \tilde{r}_{sk}\right\rbrace(\omega)=\mathcal{F}\left\lbrace \tilde{p}_{sk}\right\rbrace(\omega)$ in the transformed domain.
By applying the Fourier anti-transform and moving from the Heisenberg to the Schrödinger picture, we obtain
\begin{equation}\fl
    {\Theta}_{sk}=\frac{1}{2\pi}e^{-\frac{i\mathcal{H}_0t}{\hbar}}\int_{-\infty}^{+\infty}\mathcal{F}\left\lbrace \tilde{p}_{sk}\right\rbrace(\omega)\frac{\tanh\left(\frac{\omega\hbar}{2k_BT}\right)}{\frac{\omega\hbar}{2k_BT}}e^{i\omega t}d\omega e^{+\frac{i\mathcal{H}_0t}{\hbar}},
    \label{thetaint}
\end{equation}
where
\begin{equation}\fl
    \mathcal{F}\left\lbrace \tilde{p}_{sk}\right\rbrace(\omega)=\int_{-\infty}^{+\infty}e^{+\frac{i\mathcal{H}_0\tau}{\hbar}}{p}_{sk}e^{-\frac{i\mathcal{H}_0\tau}{\hbar}}e^{-i\omega \tau}d\tau.
    \label{thetafourierheisen}
\end{equation}
We can substitute Eq. (\ref{thetafourierheisen}) into Eq. (\ref{thetaint}), and we get
\begin{equation}\fl
    {\Theta}_{sk}=\frac{1}{2\pi}\int_{-\infty}^{+\infty}\int_{-\infty}^{+\infty}e^{+\frac{i\mathcal{H}_0\xi}{\hbar}}{p}_{sk}e^{-\frac{i\mathcal{H}_0\xi}{\hbar}}e^{-i\omega \xi}\frac{\tanh\left(\frac{\omega\hbar}{2k_BT}\right)}{\frac{\omega\hbar}{2k_BT}}d\xi d\omega,
    \label{thetatemp}
\end{equation}
where we adopted the substitution $\xi=\tau-t$.
We can now change the order of integration and exploit the following Fourier transform/anti-transform pair (see Ref. \cite{campbell1948}, Equation No. 612.1)
\begin{eqnarray}\fl
\label{direct}
\int_{-\infty}^{+\infty}\frac{\tanh\left(\delta\omega\right)}{\delta\omega}e^{-i\omega \xi}d\omega=\frac{2}{\delta}\log\left[\coth\left(\frac{\pi\vert\xi\vert}{4\delta}\right)\right],\\
\label{inverse}
\fl\frac{1}{2\pi}\int_{-\infty}^{+\infty}\frac{2}{\delta}\log\left[\coth\left(\frac{\pi\vert\xi\vert}{4\delta}\right)\right]e^{i\omega \xi}d\xi=\frac{\tanh\left(\delta\omega\right)}{\delta\omega}.\,\,\,\,\,\,\,\,\,\,
\end{eqnarray}
We assume $\delta=\frac{\hbar}{2k_BT}$, and we obtain from Eq. (\ref{thetatemp}), through Eq. (\ref{direct}), the relation
\begin{equation}\fl
    {\Theta}_{sk}=\frac{2k_B T}{\pi\hbar}\int_{-\infty}^{+\infty}e^{+\frac{i\mathcal{H}_0\xi}{\hbar}}{p}_{sk}e^{-\frac{i\mathcal{H}_0\xi}{\hbar}}\log\left[\coth\left(\frac{\pi k_B T\vert\xi\vert}{2\hbar}\right)\right]d\xi.
\end{equation}
We then apply the change of variable $\xi=\frac{\hbar}{k_B T}\eta$, and we finally prove Eq. (\ref{integ2a}) of the main text. 
It is  seen using Eq.  (\ref{inverse}) that when the energy base is adopted, we find Eq.  (\ref{frienefind}) again.
It is also interesting to note that for free particles $\mathcal{H}_0=K_0$ (there is only kinetic energy), and thus it is easily obtained via Eq.  (\ref{inverse}), with $\omega\to 0$, that $ {\Theta}_{sk}= {p}_{sk}$.
To conclude, we develop an interesting series expansion for ${\Theta}_{sk}$. To do this, we apply the Baker-Hausdorff formula to Eq. (\ref{integ2a}). 
We define the symbol $[\mathcal{H}_0,p_{sk}]_n$ through the recursive relation  $[\mathcal{H}_0,p_{sk}]_0=p_{sk}$, and $[\mathcal{H}_0,p_{sk}]_{n+1}=[\mathcal{H}_0,[\mathcal{H}_0,p_{sk}]_n]$. 
Then we have $[\mathcal{H}_0,p_{sk}]_1=[\mathcal{H}_0,p_{sk}]$, $[\mathcal{H}_0,p_{sk}]_2=[\mathcal{H}_0,[\mathcal{H}_0,p_{sk}]]$, $[\mathcal{H}_0,p_{sk}]_3=[\mathcal{H}_0,[\mathcal{H}_0,[\mathcal{H}_0,p_{sk}]]]$, and so on.
Hence, we get
\begin{equation}\fl
    {\Theta}_{sk}=\frac{2}{\pi}\sum_{n=0}^{+\infty}\frac{i^n}{n!(k_BT)^n}[\mathcal{H}_0,p_{sk}]_nI_n,
    \label{seriest}
\end{equation}
where
\begin{equation}\fl
    I_n=\int_{-\infty}^{+\infty}
\eta^n\log\left[\coth\left(\frac{\pi}{2}\vert\eta\vert\right)\right]d\eta,\,\,\,\, n\ge 0.
\end{equation}
We observe that $I_n=0$ is $n$ is odd, and therefore we calculate the values of $I_{2n}$, $n\ge 0$.
We consider Eq. (\ref{inverse}) with $\delta=1/2$, and we find
\begin{equation}\fl
    \frac{1}{\pi}\int_{-\infty}^{+\infty}\log\left[\coth\left(\frac{\pi\vert\eta\vert}{2}\right)\right]e^{i\omega \eta}d\eta=\frac{\tanh\left(\frac{1}{2}\omega\right)}{\omega}.
\end{equation}
Here, we substitute the classical power series for the exponential 
$e^{i\omega \eta}=\sum_{k=0}^{+\infty}\frac{1}{k!}(i\omega\eta)^k$, and the following power series for the hyperbolic tangent
\begin{equation}\fl
    \tanh x=\sum_{n=1}^{+\infty}\frac{2^{2n}\left(2^{2n}-1\right)B_{2n}x^{2n-1}}{(2n)!}, \,\,\,\,\vert x\vert<\frac{\pi}{2},
\end{equation}
where $B_k$ are the Bernoulli numbers. By equalizing the coefficients of the power series, it is easily found that 
\begin{eqnarray}\fl
I_{2n}=2\pi(-1)^n\frac{2^{2n+2}-1}{(2n+1)(2n+2)}B_{2n+2},
\end{eqnarray}
for $n\ge 0$ (moreover, $I_{2n-1}=0$ for $n\ge 1$).
By substituting this result in Eq. (\ref{seriest}), we get
\begin{equation}\fl
    {\Theta}_{sk}=4\sum_{n=0}^{+\infty}\frac{1}{(k_BT)^{2n}}[\mathcal{H}_0,p_{sk}]_{2n}\frac{2^{2n+2}-1}{(2n+2)!}B_{2n+2}.
    \label{seriestfin}
\end{equation}
Since $B_2=1/6$, $B_4=-1/30$, and $B_6=1/42$, we have the expansion given in Eq. (\ref{threeterms}) of the main text.
 {
\section{Friction operator with degenerate energy spectra}
\label{appd}
We suppose to consider a system with a degenerate discrete spectrum, described by the eigenvalues/eigenvectors equation  $\mathcal{H}_0\varphi_{nk}=E_n\varphi_{nk}$, where $n$ is the principal quantum number  enumerating the distinct energy levels, and $k$ represents the degeneration. 
Thus, for each value of $n$ there are multiple eigenfunctions identified by the $k$ index. Every eigenspace can be orthogonalized, and thus we can write $\langle \varphi_{nk}\vert\varphi_{mh} \rangle=\delta_{nm}\delta_{kh} $.
Such a degenerate discrete basis can be used to represent any element of the Hilbert space of wave functions. 
Indeed, given the element $\Psi$, we can write the development $\Psi=a_{nk}\varphi_{nk}$ (we adopt the Einstein summation notation for both $n$ and $k$), with coordinates $a_{nk}=\langle \varphi_{nk} \vert\Psi\rangle\in\mathbb{C}$. 
If we have a linear operator $\tau=A\Psi$, where $\tau$ belongs to the same Hilbert space as $\Psi$, we can write $\tau=Aa_{nk}\varphi_{nk}=a_{nk}A\varphi_{nk}$. 
Hence, the coordinates of $\tau$ can be obtained as $\tau_{mh}=\langle \varphi_{mh}\vert\tau \rangle=\langle \varphi_{mh} \vert A\varphi_{nk} \rangle a_{nk}$. Each operator $A$ can be therefore represented in the degenerate basis by means of the object $A_{mh,nk}=\langle \varphi_{mh} \vert A\varphi_{nk} \rangle $ with four indices. 
It is true for coordinate operators, momentum operators and also for the density matrix. We search for the representation of the friction operator in this degenerate case. 
We start by observing that the representation of the Hamiltonian operator is found as $\mathcal{H}_{0,mh,nk}=\langle \varphi_{mh} \vert \mathcal{H}_0\varphi_{nk} \rangle =\langle \varphi_{mh} \vert E_n\varphi_{nk} \rangle= E_n\langle \varphi_{mh} \vert\varphi_{nk} \rangle= E_n\delta_{nm}\delta_{kh}$.  
In the same basis, the exponential  operator $e^{-\frac{\mathcal{H}_0}{k_BT}}$ can be represented by means of the expression $\left(e^{-\frac{\mathcal{H}_0}{k_BT}}\right)_{mh,nk}=e^{-\frac{E_n}{k_BT}}\delta_{nm}\delta_{kh}$. 
As before, to simplify the notation, we introduce the quantity $e_n=e^{-\frac{E_n}{k_BT}}>0$. 
Therefore we can write that $\left(e^{-\frac{\mathcal{H}_0}{k_BT}}\right)_{mh,nk}=e_n\delta_{nm}\delta_{kh}$. Moreover, we also obtain the useful representation $\left(e^{-\xi e^{-\frac{\mathcal{H}_0}{k_BT}}}\right)_{mh,xy}=e^{-\xi e_m}\delta_{mx}\delta_{hy}$. 
The central operator in Eq. (\ref{integ1a}), $[r_{st},e^{-\frac{\mathcal{H}_0}{k_BT}}]$, is composed of the following elements
\begin{eqnarray}\fl
\nonumber
[r_{st},e^{-\frac{\mathcal{H}_0}{k_BT}}]_{xy,ij}&=&r_{st,xy,ab}\left(e^{-\frac{\mathcal{H}_0}{k_BT}}\right)_{ab,ij}-\left(e^{-\frac{\mathcal{H}_0}{k_BT}}\right)_{xy,ab}r_{st,ab,ij}\\
\nonumber\fl &=&r_{st,xy,ab}e_a\delta_{ai}\delta_{bj}-e_x\delta_{xa}\delta_{yb}r_{st,ab,ij}
\\\fl &=&r_{st,xy,ij}e_i-e_xr_{st,xy,ij}=(e_i-e_x)r_{st,xy,ij},
\end{eqnarray}
and therefore the representation of the friction operator assumes the form
\begin{eqnarray}\fl
\nonumber
    \Theta_{st,mh,nk}&=&i\frac{2 m_t k_B T }{\hbar}\int_0^{+\infty}
\left(e^{-\xi e^{-\frac{\mathcal{H}_0}{k_BT}}}\right)_{mh,xy}[r_{st},e^{-\frac{\mathcal{H}_0}{k_BT}}]_{xy,ij}\left(e^{-\xi e^{-\frac{\mathcal{H}_0}{k_BT}}}\right)_{ij,nk}d\xi\\
\nonumber
\fl&=&i\frac{2 m_t k_B T }{\hbar}\int_0^{+\infty}
e^{-\xi e_m}\delta_{mx}\delta_{hy}(e_i-e_x)r_{st,xy,ij}e^{-\xi e_i}\delta_{in}\delta_{jk}d\xi\\
\nonumber
\fl&=&i\frac{2 m_t k_B T }{\hbar}\int_0^{+\infty}
e^{-\xi e_m}(e_n-e_m)r_{st,mh,nk}e^{-\xi e_n}d\xi\\
\fl&=&i\frac{2 m_t k_B T }{\hbar}
\frac{e_n-e_m}{e_n+e_m}r_{st,mh,nk}.
\end{eqnarray}
This expression represents the friction operator on an arbitrary degenerate basis and can be used for developing and solving the master equation  in the degenerate case we are describing. 
Recall that in such a case all operators will have four indices and thus also the principal unknown given by the density matrix.
}

\end{document}